\documentclass[10pt]{iopart}
\usepackage{iopams}  
\usepackage{epsfig,xspace}
\begin{document}
     \newcommand{\pathstrange}{/home/rafelski/figure/}
     \newcommand{\pathmarc}{/usr1/rafelski/figure/}
     \newcommand{\pathbook}{/usr1/rafelski/book/figures/}
     \newcommand{\pathtransp}{/usr1/rafelski/paper/transp/}
    \newcommand{\pathlapbook}{/home/rafelski/book/qgp/figures/}
     \newcommand{\pathletes}{/users/lpthe/jletes/bookraf/figures/}
     \newcommand{\pathjussieu}{/users/visit/rafelski/figures/}
     \newcommand{\pathnow}{}
\renewcommand{\topfraction}{.99}\renewcommand{\textfraction}{.0}
\newcommand{\nc}{\newcommand}
\newcommand{\ds}{\displaystyle}
\newcommand{\ts}{\textstyle}
\newcommand{\rf}[1]{figure\ \ref{#1}}
\newcommand{\req}[1]{equation\ (\ref{#1})}
\newcommand{\rt}[1]{table\ \ref{#1}}
\nc{\eps}{\varepsilon}
\nc{\la}{\lambda}
\nc{\ga}{\gamma}
\nc{\Ga}{\Gamma}
\nc{\de}{\delta}
\nc{\De}{\Delta}
\nc{\al}{\alpha}
\nc{\be}{\beta}
\nc{\ra}{\rightarrow}
\nc{\llra}{\longleftrightarrow}
\nc{\Llra}{\Longleftrightarrow}
\nc{\Ra}{\Rightarrow}
\nc{\nf}{\infty}
\nc{\Lra}{\Longrightarrow}
\nc{\beq}{\begin{equation}}
\nc{\eeq}{\end{equation}}
\nc{\beqa}{\begin{eqnarray}}
\nc{\eeqa}{\end{eqnarray}}
\nc{\bfi}{\begin{figure}}
\nc{\efi}{\end{figure}}
\def\agev{{$A$~GeV}\xspace}
\def\AGeV{\agev}

\title[Strangeness and QGP]{Strangeness and Quark Gluon Plasma
}

\author{Johann Rafelski\dag \ and Jean Letessier\ddag\dag} 
\address{\dag\ Department of Physics,
           University of Arizona, Tucson, AZ 85721}
\address{\ddag\ Laboratoire de Physique 
                Th\'eorique et Hautes Energies\\
    \ \   Universit\'e Paris 7, 2 place Jussieu, F--75251 Cedex 05.}

\begin{abstract}
A brief summary of strangeness mile stones is followed by 
a chemical non-equilibrium statistical hadronization
analysis of strangeness results at SPS and RHIC. 
Strange particle production in $AA$ interactions 
at $\sqrt{s_{NN}}\ge 8.6$ GeV  can be understood 
consistently as originating from the deconfined 
quark--gluon plasma in a sudden hadronization process.
Onset of QGP formation as function of energy is placed
in the beam energy interval 10--30$A$ GeV/c. Strangeness
anomalies at LHC are described.
\end{abstract}


\submitto{\JPG}


\section{Introduction}
\label{intro}
We consider, in a chemical analysis of yields, 
the strange hadron production in
 A--A collisions at the SPS  and RHIC. We find that  
these results can be well understood and 
consistently described within the  Fermi type, chemical near-equilibrium, 
statistical  hadronization model~\cite{Let99,Let00b,Let02gp}.
We find that the deviation from equilibrium conditions 
are most pronounced at RHIC ($\sqrt{s_{NN}}=130$ GeV, 
RHIC$_{130}$)~\cite{Raf01kc,Raf02ga},
 with the conditions of hadronization strongly deviating from 
chemical equilibrium. The present analysis and 
reanalysis of the results obtained in the energy range 
$19.4\le \sqrt{s_{NN}}\le 8.75$ GeV by 
NA49~\cite{App98,Afa01qj,Afa02ub,Afa02he,Ale03gy,Hoh02pb,Gaz03} and 
WA97/NA57~\cite{And99,Ant99a,Ant00e,Ant02we} experiments
is confirming these features characteristic of the formation of
a new state of matter at SPS. 

The strongest difference between SPS and RHIC collision results is
the order of magnitude enhancement in strange particle yield
per participant, this result is not new but has found so 
far little attention~\cite{Raf01kc,Raf02ga}. 
The strangeness yield rise
increases compared to SPS~\cite{Let00b,Let02gp} in a manner
which is  more spectacular than the increase in the total hadron 
multiplicity. This is an expression of the increase of the 
excess of strangeness (greater chemical nonequilibrium) at RHIC as 
compared to SPS. We establish  this increase in strangeness 
yield with precision, which requires use of an appropriate
analysis tool of hadron multiplicities, which allows reliable 
and precise description of the unobserved particle phase space 
in domains   not yet experimentally  accessible. 

At RHIC$_{130}$, each baryon participant 
leads to formation of nearly
9 strange quark pairs in Au--Au reactions. 
We show  in section~\ref{SPSRHIC} that 
this high yield is reached gradually as function 
of the available fireball thermal energy. This  is 
confirming that this high yield of strangeness is
result of the same physical process in all
reaction energies considered here, beginning with 
$\sqrt{s_{NN}}=8.75$ GeV (corresponding
to a 40$A$ GeV beam on laboratory stationary target).
We also present in section~\ref{SPSRHIC}
how the diverse properties of 
the hadronic fireball change as function of energy. 
The introduction of chemical
nonequilibrium leads to  a rather low 
hadronization temperature which we interpret in section \ref{HQGP}.
where we discuss in depth at which energy domain the
onset of QGP may be occurring.

Before we turn to this part of our report a brief introduction
to `strangeness' the important mile stones  
of 10 years of experimental research are presented
in the following section~\ref{tool}. 
This is followed in section \ref{sprod} by a 
 discussion of the mechanism 
of QCD based production of strangeness 
in a quark--gluon plasma
(QGP). In particular we show that the gluon fusion
processes in QGP is capable to populate within the
reaction time the strange quark 
phase space~\cite{Raf82b,Koc86b,Let96,Raf99a}, 
up to, and even above the chemical equilibrium 
yield in the QGP phase~\cite{Raf99a}.

The other QCD based process, light quark--antiquark 
fusion into strange quark pair, has been earlier shown 
to be too slow  for strangeness equilibration 
in the QGP~\cite{Bir82,Bir82b}. Thus the discovery of
the dominance of the 
gluon fusion process has been the  key stepping
stone in establishing strangeness as an observable of QGP.
The computation presented in section~\ref{sprod} is 
assuming early thermalization,  and following rapid collective 
evolution. This scenario  is  today  viewed as a very 
likely reaction sequence, seen the behavior of the $v_2$ flow
parameter, especially so that it applies also to 
strange hadrons~\cite{Adl02pb}.

A very important aspect of the study of strange particles
produced by QGP is the final state hadron production process. 
We present the essential elements of 
statistical hadronization in section~\ref{stathad}. These 
principles are employed  in section~\ref{SPSRHIC}, in analysis of the 
RHIC and SPS results, and as noted a synthesis of our analysis and 
their impact for the search of the onset of QGP formation 
is offered in section~\ref{HQGP}.
In the final section \ref{lhc} we argue that there is a very 
interesting future for strangeness in the LHC energy domain.

\section{Strangeness --- a popular QGP diagnostic tool}
\label{tool}
 There are  several different  strange particles in nature
allowing us to  study
several complementary  physics questions. Denoting here 
light flavors with $q=u,d$
we consider  the stable particles:
\[
\fl \hspace*{1cm}
\phi(s\bar s),\ {\rm K}(q\bar s),\ \overline{\rm K}(\bar q s),
\ \Lambda(qqs),\ \overline{\Lambda}(\bar q\bar q\bar s),
\ \Xi(qss), \ \overline{\Xi}(\bar q\bar s\bar s),
\ \Omega(sss), \ \overline{\Omega}(\bar s\bar s\bar s).
\]
Moreover, the study of strange hadron resonances such as
${\rm K}^*,\ \Sigma^*,\ \Lambda^*$ is a very promising avenue of 
forthcoming research, leading to direct determination of hadronization 
conditions. 

 Many strange hadrons are subject to a  self analyzing decay
within  several centimeters  from the point of production, as is illustrated
in \rf{VSIG} for the cascading $\Xi^-$-decay. Thus one tracking device,
typically today a TPC combined with silicon vertex detector allows
the measurement of the many particle 
yields of interest. The accessibility of strange
hadron distributions has lead, in recent years, to an explosion of 
both experimental and theoretical interest. Moreover,  
production rates of hadronic particles and hence   
statistical significance even for low intensity beam 
 is usually high.

\begin{figure}[t]
\centerline{\hspace*{1cm}
\psfig{width=8cm,figure=\pathnow 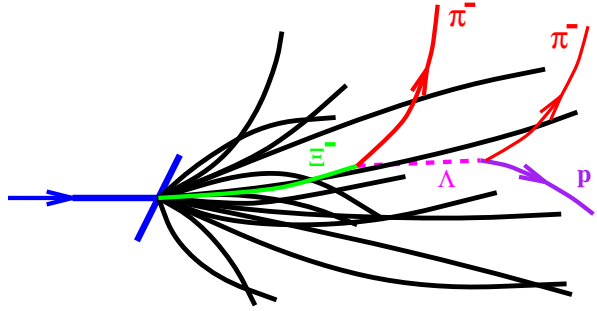}
	}
\vspace*{-0.5cm}
\caption{\label{VSIG}
Illustration of the $\Xi^-$-production by an 
incoming, e.g., SPS Pb-beam,  on a heavy laboratory target, and its subsequent 
decay. Dashed line depicts the invisible neutral $\Lambda$, emerging 
from the decay kink $\Xi^-\to \pi^-\Lambda$, and ending in the decay `V' 
of $\Lambda\to p+\pi^-$. 
}
\end{figure}

Multistrange hadrons  can be formed in hadronization of
 QGP by  `cross talk'  between quarks made
in disjoint microscopic reactions~\cite{Raf82a}. An illustration of this 
process is presented in \rf{JRSTRANGEPROD}: the inserts show 
gluon fusion processes $gg\to s\bar s$~\cite{Raf82b}, which
establish ample supply of strange quark pairs 
in the early stages of QGP evolution. 
In the ensuing hadronization process, quark 
recombination leads to emergence of particles such
as here shown $\Xi(sqq),\ \overline{\Omega}(\bar s\bar s\bar s)$, 
which otherwise could only very 
rarely be produced, considering that several 
rarely occurring processes would 
have to coincide. Enhancement of such particles
is for this reason indicative of deconfinement~\cite{Raf82b,Raf87b}. 
This mechanism leads to the expectation that the enhancement of strange
antibaryons is progressing with strangeness content,
as was recognized and discussed qualitatively 
already 20 years ago, see e.g., section 4 in~\cite{Raf84b}.

\begin{figure}[t]
\centerline{\epsfig{width=8cm,angle=-90,figure=\pathnow 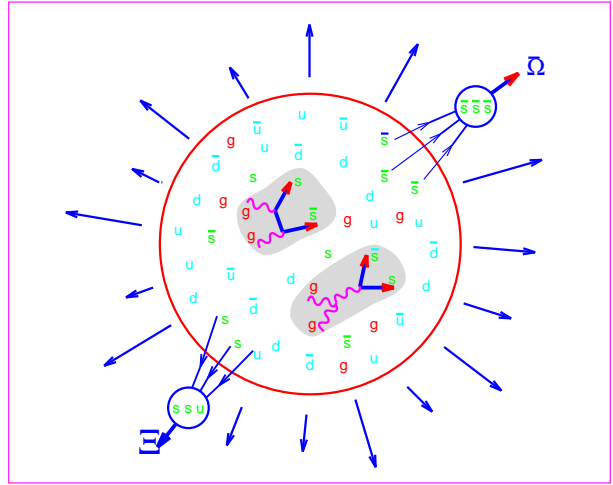}}
\vspace*{-1.5cm}
\caption{
Illustration of the cross-talk two step mechanism of strange hadron formation 
from QGP: inserts show gluon fusion into
 strangeness, followed by QGP recombinant hadronization.
\label{JRSTRANGEPROD}
}
\end{figure}

Since the production of strangeness occurs predominantly 
in thermal gluon fusion $gg\to s\bar s$, 
see section \ref{sprod}, the overabundant presence of 
strangeness in high energy relativistic reactions can be on theoretical
grounds  linked to presence of thermal gluons, and thus to QGP. 
Beyond the  dominant processes $gg\to s\bar s$  there is also, 
at 10--15\% of the total rate, a contribution from  $q\bar q\to s\bar s$.
This rate alone would however not suffice to equilibrate strangeness in
rapidly disintegrating QGP phase formed in relativistic heavy ion 
collisions~\cite{Bir82,Bir82b}.

The  remarkable  coincidence of scales
involving the strange quark mass $m_s$, and the critical temperature of
hadronic phase transition $T_c$,
\begin{equation}
m_s\simeq T_c  \quad \mbox{which\ implies}\quad  \tau_s \simeq \tau_{\rm QGP},
\end{equation}
allows the  $gg\to s\bar s$ production process to act as a clock for the
collision reaction.
Another interesting feature is that, in a baryon rich environment,
we also have $\bar s>\bar q$~\cite{Raf82a}, which applies
 when the quark  chemical potential is greater than the 
mass of the strange quark. In this condition
a   strange  antibaryon  enhancement  is 
expected, comparing to the nonstrange
antibaryons, including the anomaly $\overline\Lambda/\bar p>1$. This should 
be the case at the lower energy limit of SPS. 

In this line of thought, 
even though there is a near $u,d,s$ 
flavor symmetry at  RHIC, we  also expect 
(anti)hyperon dominance of (anti)baryons in this condition. Here,
 the mechanism 
is slightly different~\cite{Raf99a}: each baryon or antibaryon contains three
quarks or, respectively, antiquarks, and it is hard to find in the 
flavor symmetric pool
of quarks (and antiquarks) three non-strange quarks (or antiquarks), 
in fact the statistical probability for assembly of three nonstrange 
quarks is  1/3. Thus 2/3 of all produced
baryons, or respectively antibaryons,  will in limit of flavor symmetry
carry strangeness,
and this feature of course remains true in the future LHC environment.

Multistrange anti-hyperon abundance ratios were studied extensively 
and  we show  in \rf{JRALLAB03} the three ratios:
\[
\fl \hspace*{1cm}
{\bar s\bar s\bar d\over\bar s\bar u\bar d}=
\frac{\overline{\Xi^-}+0.5\,\overline{\Xi^*(1530)}}
{\overline{\Lambda}+\overline{\Sigma^0}+0.92\,\overline{\Sigma^*(1385)}},\qquad
{ssd\over sud}=\frac{\Xi^-+0.5\,\Xi^*(1530)}
{\Lambda+\Sigma^0+0.92\,\Sigma^*(1385)},
\]
\[
\fl \hspace*{1cm}
{\bar s\bar s\bar d+ssd\over\bar s\bar u\bar d+ sud}=
\frac{\overline{\Xi^-}+0.5\,\overline{\Xi*(1530)}+\Xi^-+0.5\,\Xi^*(1530)}
{\overline{\Lambda}+\overline{\Sigma^0}+0.92\,\overline{\Sigma^*(1385)}
+\Lambda+\Sigma^0+0.92\,\Sigma^*(1385)}.
\]
Above,  
we have explicitly identified the 
main resonance contributions, accompanied by~the appropriate 
branching ratio. Given their 
large spin-isospin quantum numbers, these resonances 
despite their greater mass are 
as important contributors to the final yields, as is the 
ground state. For the $\Lambda$, we also
note the presence of inseparable 
$\Sigma^0\to \Lambda+\gamma$ contribution. 

\begin{figure}[ht]
\vskip-0.cm
\hspace*{2cm}\psfig{width=10cm,height=10cm,figure=\pathnow 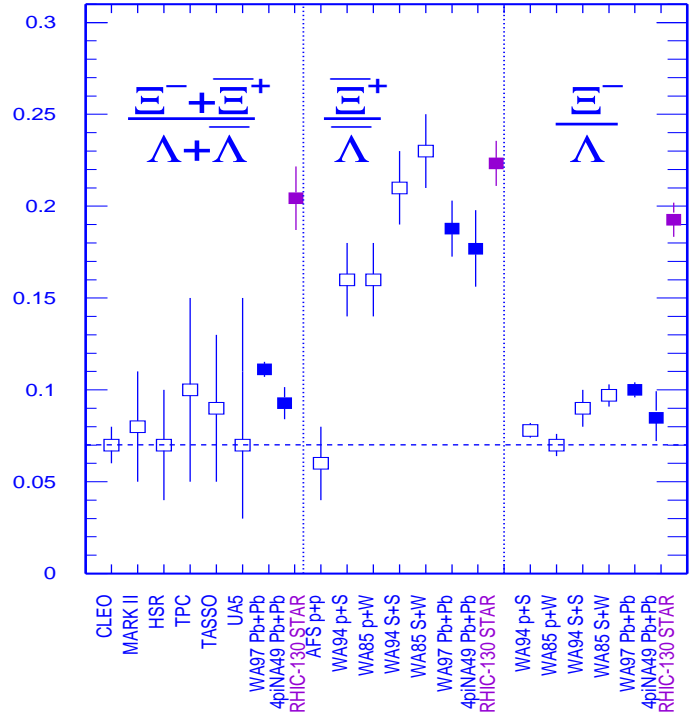}
\vskip-0.6cm
\caption{
Ratios of hyperon abundances produced in high energy interactions. 
Recent experimental results
 from~\protect\cite{Ake84,Aln85,Ale90,Ant00e,App98,Afa02he,Hua02rx,Cas02xv}
\label{JRALLAB03}.
}
\end{figure}

It is practically impossible 
to separate decay contributions which dilute  the 
intrinsic ratios $ssd/sud$, $\bar s\bar s\bar d/\bar s\bar u\bar d$ 
down to about 0.23 for the case of RHIC~\cite{Cas02xv}.  However, we see  a clear
enhancement (factor 3) over the $p$--$p$, $p$--$\bar p$ background for all A--A interactions 
here considered. Among these,
 the RHIC result is of most profound consequence as we shall 
address below, as it implies large excess of strangeness 
yield (compare~\rf{JRPLGSGQT3}). 
The enhancement of the antibaryon rations seen in the $p$--A ratio 
(middle field of \rf{JRALLAB03}) is {\em not} seen for baryons (see right field of
 \rf{JRALLAB03}). A simulated $p$--A enhancement
can arise from the annihilation of  antihyperons produced by the reacting 
matter in the  in surrounding spectator nuclear matter. Namely, 
the annihilation  cross section for singly strange hyperons is notably 
 larger than for double strange hyperons. In consequence the 
$\overline\Lambda$ are more depleted than $\overline\Xi$. 

To avoid the semblance of $p$--A enhancement process it is important 
to consider how the yield of individual hadrons behaves compared to 
expectations based on a cascade of N--N interactions.
This view on the strange hadron enhancement has been 
presented by the CERN-SPS WA97 collaboration for 158$A$ GeV Pb collisions with 
laboratory Pb target, and these results have been confirmed and extended  in the
more recent NA57 measurements. In \rf{JRSenhsys1}, we show the rise
of the enhancement with the strangeness content, with $p$--Be reaction system
used as reference. These results show yields per participant, where 
 a participant is  understood to be an inelastically reacting `wounded' nucleon.
These results confirm the QGP prediction of enhancement growth 
with a)~strangeness b)~antiquark content.

\begin{figure}[ht]
\vspace{-1.2cm}
\centerline{\hspace*{1.5cm}\psfig{width=12cm,figure=\pathnow 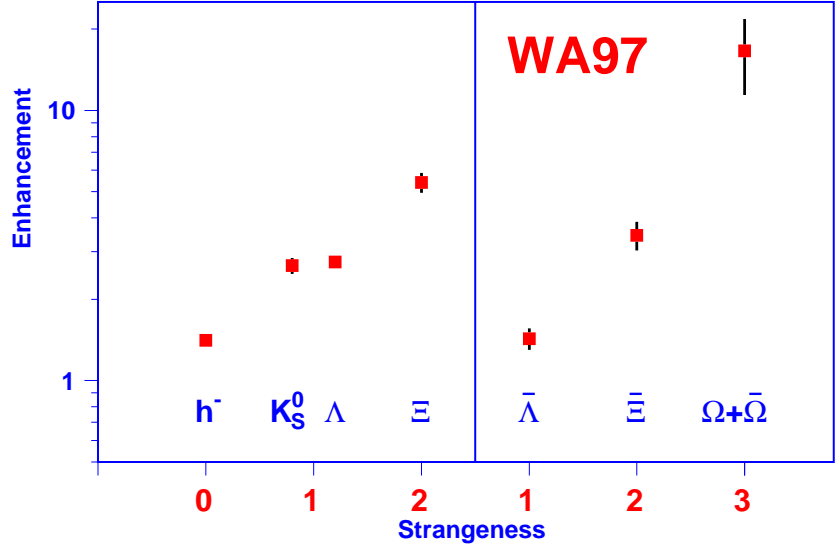}}
\vskip-0.4cm
\caption{\label{JRSenhsys1}
Abundance enhancement with respect
to the yield in  p--Be  collisions, scaled 
 with the number of `wounded' nucleons, 
results of the WA97 collaboration~\protect\cite{And99}.
}
\end{figure}

\begin{figure}[ht]
\vspace{-0.5cm}
\centerline{
\hspace*{-1.2cm}\psfig{width=8cm,figure=\pathnow 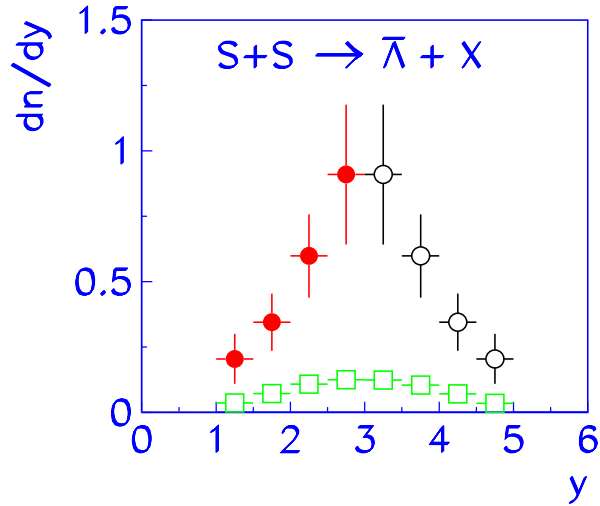}}
\vspace{-.3cm}
\caption{\label{JRNA35antLambda}
$\overline\Lambda$, in NA35II 200$A$ GeV S--S interactions~\protect\cite{Alb94}. 
Background (squares) from multiplicity scaled N--N reactions
 forward  (open circles) points: reflection at the CM rapidity.
}
\end{figure}

Before these studies of strangeness enhancement were
carried out within the central rapidity region by 
NA57 and WA97 experiments, it has
been shown that  central rapidity is indeed
the kinematic domain where the new physics is occurring.
The NA35II experiment obtained  $\overline\Lambda$ yield for  200$A$ GeV 
Sulfur projectile  on laboratory Sulfur target~\cite{Alb94,Alb96}. 
As can be seen in \rf{JRNA35antLambda},
the antibaryon yields are localized in the  central CM (center of momentum) 
 rapidity region. The background  $p$--$p$ production result, shown by 
squares is obtained scaling up the  yield by the number of nucleon participants. 
We see that the central rapidity is in fact the domain of the large 
enhancement as would be expected in a reaction picture in which a 
fireball of hot matter is formed.

Strangeness yield as function of rapidity is obtained  evaluating 
$\langle s+\bar s\rangle =1.6\Lambda+1.6\overline\Lambda+4\mbox{K}_S$, and
is seen in \rf{JRSS}. Again, the measured  result is reflected at 
CM rapidity and the squares provide a hadron multiplicity
 scaled $p$--$p$ result as basis.  Thus, the  enhancement, seen in \rf{JRSS},
is expressing how much {\em faster} than nonstrange hadron 
strangeness increases in S--S interactions compared to $p$--$p$. 

\begin{figure}[t]
\vspace{-0.5cm}
\centerline{
\hspace*{0.5cm}
\psfig{width=8cm,figure=\pathnow 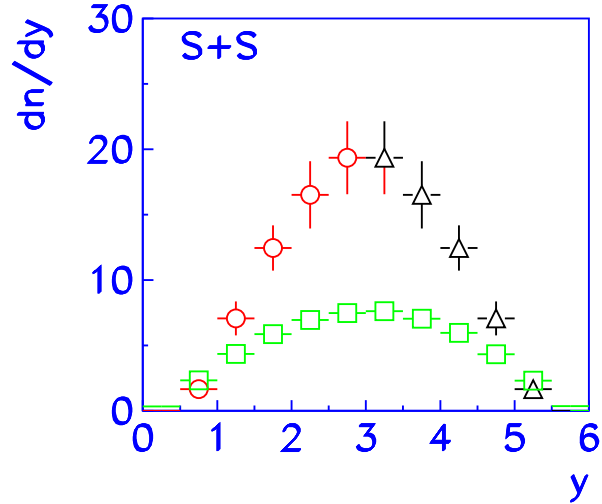}}
\vspace{-.3cm}
\caption{\label{JRSS}
CM-rapidity reflected 1.6$\Lambda+1.6\overline\Lambda+4\mbox{K}_S$
200$A$ GeV S--S multiplicity scaled 200$A$ GeV $p$--$p$ 
results drawn from NA35II experiment~\protect\cite{Alb96}.
}
\end{figure}
One would expect naively, in
such a comparison, a reduction rather than an enhancement, since in
a cascade of conventional hadronic reactions, it would be easier to 
produce less massive pions, than strangeness, remembering that cascading
hadrons degrade in energy. There can be no doubt seeing this 
result that even in the relatively small S--S reaction system there
is a new physical mechanism of strangeness production at work. In absolute 
number, the yield of strangeness, seen in \rf{JRSS}, is rather 
large, above 25 $s\bar s$ pairs. Our predictions that this could be
happening should  QGP phase be formed  were classified 20 years ago at the LBL 
Heavy Ions Studies among `exotica'.

\begin{figure}[t]
\vspace{0.4cm}
\hspace*{2.5cm}\psfig{width=8cm,figure=\pathnow 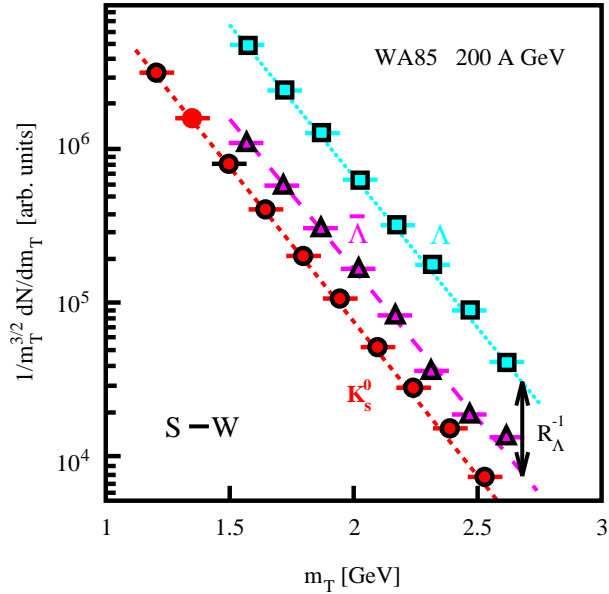}
\caption{\label{JRLALABARKS}
Central rapidity spectra of single strange K$_{\rm S}$,
$\Lambda$ and $\overline\Lambda$ reported 
by the experiment WA85~\cite{Dib95}. Antibaryon to baryon ratio  
$R_\Lambda\equiv\overline\Lambda/\Lambda$ 
is independent of $m_\bot$.
}
\end{figure}

Aside of particle yields, an important further key piece of
physical evidence is  the shape of transverse mass 
$ m_\bot=\sqrt{m^2+p_\bot^2}$ spectra. The experimental results 
cannot be  understood within a conventional hadron cascade reaction picture. 
When the (central rapidity) $m_\bot$ spectral  distributions 
are fitted to an exponential shape, one 
finds that the inverse $m_\bot$  slopes show a rather precise 
baryon-antibaryon universality, even though at CERN-SPS 
significant baryon number asymmetry prevails. This was first
discovered in S-induced collisions
 by WA85 experiment~\cite{Dib95}, see~\rf{JRLALABARKS}.
This is confirmed by the WA97 experiment
 in Pb--Pb Interactions~\cite{Ant00}, see \rf{WA97QALLHYP}, with the 
slopes given in table~\ref{Tbot}. The inverse  slopes of 
the baryon and antibaryon spectra are to a very 
great precision (1\%) the same.
Also, within the error, the slopes of $\Lambda$ and $\Xi$ are the same.
Results  of the experiment NA57 
indicate that the more precisely measured $\Omega$ and
$\overline\Omega$ spectra are also yielding the same $T_\bot$. We 
understand the kaon-hyperon slope difference  to be result of
the collective  explosive flow~\cite{Bea97}, and the different 
range of $m_\bot$  considered evaluating the $m_\bot$  slopes.

\begin{figure}[t]
\vspace*{-8.8cm}
\hspace*{4cm}
\psfig{height=17cm,width=22.5cm,figure=\pathnow 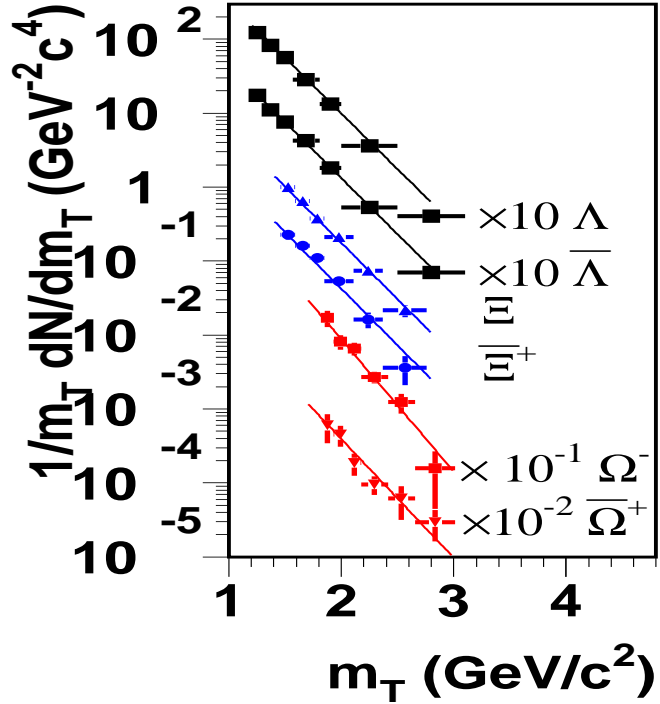}
\vspace*{1cm}
\caption{Central rapidity spectra of 
strange baryons $\Lambda(sud),\,\Xi^-(ssd),\, \Omega(sss)$ and their 
antibaryons $\overline\Lambda,\ \overline\Xi,\ \overline\Omega$ reported 
by the experiment WA97~\cite{Ant00}. 
\label{WA97QALLHYP}}
\end{figure}

These $m_\bot$-slope results, found in different reaction
systems, cannot be accidental, also since a   different pattern of
behavior is clearly seen in $p$--$p$ and $p$--A reactions. 
The identity of baryon and antibaryon $m_\bot$ slopes observed
in A--A reactions prove that 
strange baryons and antibaryons are produced by the same
mechanism in nuclear collisions, 
and do not suffer rescattering in a dense baryon-rich 
medium after their production. A systematic experimental~\cite{Ant00} and 
theoretical~\cite{Tor01} study of the $m_\bot$ spectra for the 
central rapidity Pb--Pb reactions at projectile energy $158A$ GeV 
as function of reaction centrality 
confirms that Kaons and strange hyperons and antihyperons 
are born simultaneously and do not undergo rescattering. 
To the best of our 
understanding the only physical mechanism  which can lead to
this result is a QGP fireball breakup into free streaming 
hadrons, as would be expected in 
sudden hadronization of a QGP fireball~\cite{Raf91a,Cso94,Cse95,Raf00}.

\begin{table}[t]
\caption{\label{Tbot} The inverse slope $T_\bot$  of $m_\bot$ spectra
 seen in figure \protect\ref{WA97QALLHYP}, Pb--Pb interactions
at 158$A$ GeV, WA97 experiment~\cite{Ant00}.}
 \begin{indented}\item[] 
\begin{tabular}{l|c|cc|cc|c}
\br
Particle   &${\rm K}_{\rm S}$ & $\Lambda$ & $\overline\Lambda$ & $\Xi$ & $\overline\Xi$ & $ \Omega+\overline\Omega$ \\
\mr
$T_{\bot}$\,[MeV]& 230$\pm$2 & 289$\pm$3& 287$\pm$4 & 286$\pm$9& 284$\pm$17& 251$\pm$19\\
\br
\end{tabular}
 \end{indented}
 \end{table}

In \rf{JRmt}, we show a survey of the SPS-158$A$ GeV transverse slopes.
The inverse slope parameter  $T$ is related to the  intrinsic 
thermal parameters $T_\mathrm{tf}$  in the source and 
local flow velocity $v_\mathrm{tf}$, for $m,p_\bot\gg T$
by the usual  Doppler formula : 
\begin{equation}
T\simeq{ {1+\vec n \cdot  \vec v_\mathrm{tf}}\over 
     {\sqrt{1-\vec v^{\,2}_\mathrm{tf}}}} T_\mathrm{tf}
\to
\sqrt{1+v_\mathrm{tf}\over 1-v_\mathrm{tf}} T_\mathrm{tf}.
\end{equation}
\begin{figure}[t]
\vspace*{0cm}
\hspace*{2.5cm}\epsfig{width=9cm,figure=\pathnow 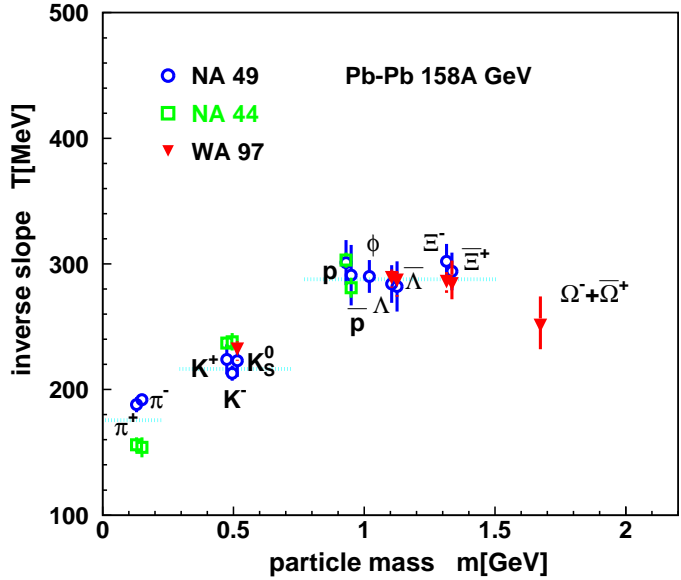}
\vspace*{0cm}
\caption{\label{JRmt}
Inverse slopes of particle spectra obtained for Pb--Pb interactions
at projectile energy 158$A$ GeV.}
\end{figure}

\section{Kinetic description of strangeness production in QGP}
\label{sprod}
We view a QGP fireball as
consisting of  quarks and gluons constrained by 
the external `frozen color vacuum'. Within the deconfined
domain these particles can move and undergo collisions, and 
reactions, within 
a volume much larger than the usual nucleon size. The 
reactions occur at energies characterized by thermal equilibrium 
and thus the kinetic process like this is often called
`thermal production'  in order to distinguish these soft 
reactions  from hard
kinetic processes occurring prior to thermalization of
the incoming particle momenta. These first collision 
processes play a decisive role in providing abundance
of charm, and heavier flavors, in quark--gluon plasma.

However, `thermal production'  can mean something entirely 
different in other related work. Namely, this term has
 also been used to describe the statistical 
yield of particles in chemical  equilibrium. Both usages 
coincide only when the dynamics of the reaction process allows for the 
kinetic processes to establish  chemical equilibrium. 
This, in general, will not be the case in heavy ion 
collisions at relativistic energies.

The generic angle averaged cross sections for (heavy) flavor $s,\, \bar s$
production processes
$
g+g\to s+\bar s $ and $ q+\bar q\to s+\bar s\,,
$
in lowest order are well known:
\begin{equation}
\fl
\bar\sigma_{gg\to s\bar s}(s) =
   {2\pi\alpha_{\rm s}^2\over 3s} \left[
\left( 1 + {4m_{\rm s}^2\over s} + {m_{\rm s}^4\over s^2} \right)
{\rm tanh}^{-1}W(s)-\left({7\over 8} + {31m_{\rm s}^2\over
8s}\right) W(s) \right]\,,
\end{equation}
\begin{equation}
\fl
\bar\sigma_{q\bar q\to s\bar s}(s)=
   {8\pi\alpha_{\rm s}^2\over 27s}
   \left(1+ {2m_{\rm s}^2\over s} \right) W(s)\,.\qquad W(s) 
= \sqrt{1 - 4m_{\rm s}^2/s}
\end{equation}

\begin{figure}[tbh]
\vspace*{-4.7cm}
\hspace*{2cm}\psfig{height=14cm,width=14cm,figure=\pathnow 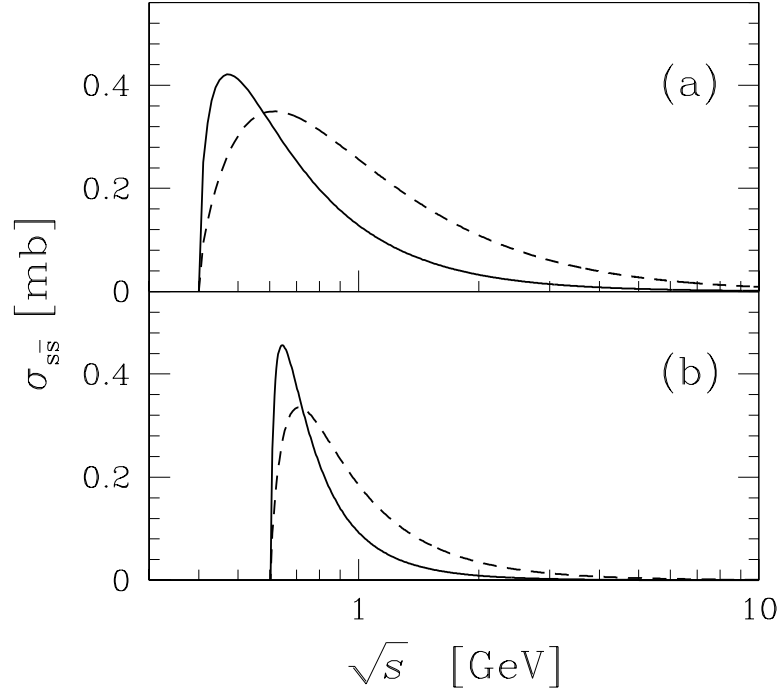}
\vspace*{-1cm}
\caption{\label{SIGMASRUN}
QGP strangeness
production cross sections~\protect\cite{Let02gp}:  
Solid lines $q\bar q\to  s\bar s$; 
dashed lines  $gg\to s\bar s$.
  a)  for fixed $\alpha_{s} = 0.6$,  $m_{\rm s}
= 200$\,MeV;
 b) for running
$\alpha_{s}(\protect\sqrt{s})$ and $m_{\rm s}(\protect\sqrt{\rm
s})$,  with $\alpha_{s(M_Z)}=0.118$. $m_s(M_Z)=90$ MeV,
$m_s(1\mbox{GeV})\simeq 2.1m_s(M_Z)\simeq 200$MeV.
}
\end{figure}

Considering that the kinetic (momentum) equilibration is faster than 
the process of strangeness production (chemical equilibration), 
we use equilibrium Fermi-Dirac or Bose particle distributions $f(\vec p_i,T)$ 
to obtain the thermally averaged strangeness production rate:
\begin{equation}
\langle \sigma v_{\rm rel} \rangle_T\equiv
\frac{\int d^3p_1\int d^3p_2  \sigma_{12} v_{12}  f(\vec p_1,T)f(\vec p_2,T)}
{\int d^3p_1\int d^3p_2 f(\vec p_1,T)f(\vec p_2,T)}\,.
\end{equation}
This leads to the (Lorentz invariant) reaction rate per unit volume and time:
\begin{equation}\label{Ainv}
 A ^{gg\to s\bar s}=
   \frac12 \rho_g^2(t)\,\langle\sigma v \rangle_T^{gg\to s\bar s}\,,\quad 
 A ^{q\bar q\to s\bar s}= \rho_{q}(t)\rho_{\bar q}(t)
\langle\sigma v \rangle_T^{q\bar q\to s\bar s}\,,
\end{equation}
\[
 A ^{s\bar s\to gg, q\bar q}= \rho_{s}(t)\,\rho_{\bar{\rm s}}(t)\,
\langle\sigma v \rangle_T^{s\bar s\to gg,q\bar q}.
\]
The factor $1/2$,  introduced here for two gluon processes, compensates  
the double-counting of identical particle pairs, arising 
since we are averaging  considering all   reacting particles
and thus each pair is counted twice. 
This invariant rate \req{Ainv} is the source term  for the  strangeness current:
\begin{equation}
\partial_\mu j^\mu_{s}\equiv  {\partial \rho_{s}\over \partial t} +
   \frac{ \partial \vec v \rho_{s}}{ \partial \vec x}=
    A ^{gg\to s\bar s}+ A ^{q\bar q\to s\bar s}-
     A ^{s\bar s\to gg, q\bar q}.
\end{equation}

In the local rest frame of reference  ($\vec v =0$) and ignoring the collective flow 
of matter we have:
\begin{equation}
\fl
{d \rho_{s}\over d t}={d \rho_{\bar s}\over d t} 
=
 \frac12 \rho_g^2(t)\,\langle\sigma v \rangle_T^{gg\to s\bar s}
+
 \rho_{q}(t)\rho_{\bar q}(t)
\langle\sigma v \rangle_T^{q\bar q\to s\bar s}
- \rho_{s}(t)\,\rho_{\bar{\rm s}}(t)\,
\langle\sigma v \rangle_T^{s\bar s\to gg,q\bar q}.
\end{equation}
Evolution for $s$ and $\bar s$ is identical, which allows to set 
$\rho_{s}(t)=\rho_{\bar s}(t)$. Using detailed balance to simplify, we obtain:
\begin{equation}
{d \rho_{s}\over dt}=
A\left (1-\frac{\rho^2_{s}(t)} {\rho^2_{s}(\infty)} \right)\,,\qquad
A= A ^{gg\to s\bar s}+ A ^{q\bar q\to s\bar s}.
\end{equation}
The generic solution at fixed $T$ ($\rho\propto \tanh (t/2\tau_s)$) implies that in 
all general cases there is an exponential approach to chemical equilibrium 
\begin{equation}\label{asym}
\fl
{\rho_s(t)\over \rho_s^\infty} \to 1-e^{-t/ \tau_s }
\qquad
 \tau_{s} \equiv
{1\over 2}{\rho_{s}(\infty)\over{
 \frac12 \rho_g^2(t)\,\langle\sigma v \rangle_T^{gg\to s\bar s}
+
 \rho_{q}(t)\rho_{\bar q}(t)
\langle\sigma v \rangle_T^{q\bar q\to s\bar s}
+\ldots}}\,,
\end{equation}
where the characteristic time constant $ \tau_s$ is the ratio of the 
density we are `chasing' with the rate at which the case occurs. There
could be additional strangeness formation processes as is suggested 
in \req{asym} in  denominator  for $ \tau_s$.

The strangeness chemical relaxation time is rather short, evaluation 
of the diverse expressions presented here produces the result seen
in \rf{TAUQGTSRUNPARIS}, where the large uncertainty is due to 
20\% uncertainty in the value of the strange quark mass. 
We recognize that the standard QCD cross sections 
  $\sigma^{gg\to s\bar s}_{\rm QCD}$ and particle
densities associated with a temperature $T>200$\,MeV yield 
 $\tau_s$  similar to lifespan of the plasma phase. Here, we note
that both $ \tau_s$  and the lifespan of plasma decrease with increasing
temperature of the initial state: for the plasma lifespan this is
due to more explosive outflow of matter driven by greater 
initial pressure, while for $ \tau_s$ this behavior 
is result of both the increased particle 
density and increased average reaction energy.

\begin{figure}[ht]
\vspace*{-0.5cm}
\hspace*{1.5cm}\psfig{width=11cm,figure=\pathnow 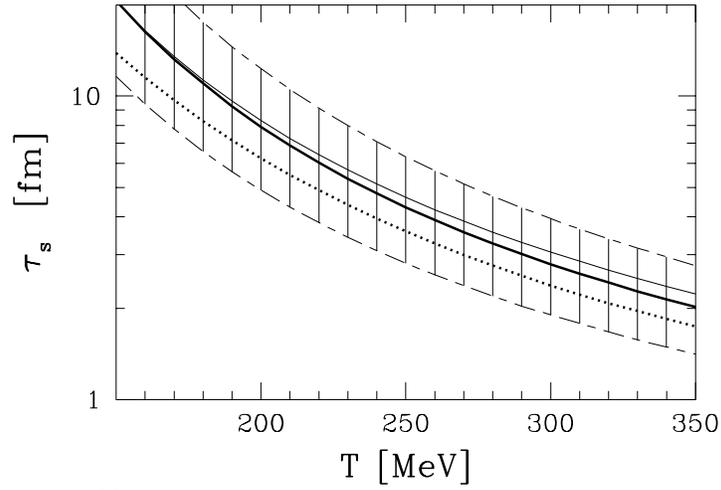}
\vspace*{-.50cm}
\caption{\label{TAUQGTSRUNPARIS}
Strangeness thermal relaxation time constant as function
of temperature. Dotted line is for fixed $\alpha_s=0.6$ and $m_s=200$\,MeV
while the solid lines are for running 
QCD parameters, see \protect\rf{SIGMASRUN}(b). Thin
solid line indicates the remaining uncertainty in  $\alpha_s$ while
the hatched area indicates the domain for 20\% variation of $m_s$.
}
\end{figure}

The temporal evolution of the strangeness density can be 
arrived at considering  entropy conserving expansion 
$T^3V=$ Const. and allowing for the dilution of the 
strange quark density by collective matter flow. Introducing the phase 
space occupancy factor $\gamma_s^{\rm QGP}$ and limiting 
ourselves to consider  Boltzmann distribution, we find 
that $\gamma_s^{\rm QGP}$ evolves in time according to:
\begin{equation}
\fl
2 \tau_{s} \frac{dT}{dt}\left({{d\gamma_{s}^{\rm QGP}}\over{dT}}+
\frac{\gamma_{s}^{\rm QGP}}{T}z\frac{K_1(z)}{K_2(z)}\right)
=1- \left(\gamma_{s}^{\rm QGP}\right)^2,\
 \gamma_s^{\rm QGP}(t)\equiv {n_s^{\rm QGP}(t)\over n_s^\infty},\
z=\frac{m_s}{T}\,.
\end{equation}
where $K_i$ are the Bessel functions. We see that the time
dependence of temperature
determines the scale at which  $\gamma_s^{\rm QGP}$  evolves.
Solutions of this equation show that one can oversaturate
the final QGP phase space, since the large initial strangeness
abundance, even if sub-equilibrium at that time, 
can exceed the equilibrium population at lower temperature~\cite{Raf99a}.

The same methods can be applied in the study of thermal charm formation. 
The relaxation time constant seen in \rf{TAUQGTCRUNPARIS2} is large, 
 indicating that charm  will not be significantly produced
in thermal parton collisions. On the other hand, charm is 
produced abundantly in first hard
parton collisions.  Benchmark values are 10 $c\bar c$ pairs
in central Au--Au at RHIC-200; and  200  $c\bar c$ pairs in central Pb--Pb LHC-6000
reactions~\cite{The03}.  
This yield is greater than the expected equilibrium 
yield at hadronization of QGP.

\begin{figure}[tbh]
\vspace*{-4.2cm}
\hspace*{1.7cm}\psfig{width=12cm,figure=\pathnow 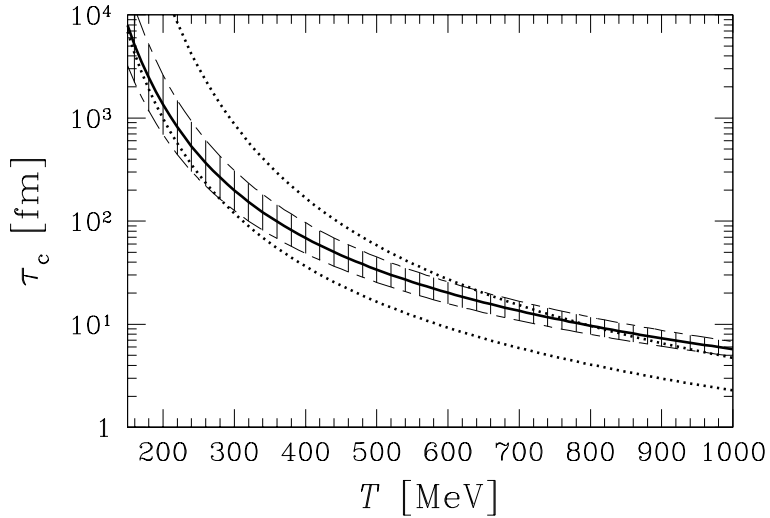}
\vspace*{-1.3cm}
\caption{\label{TAUQGTCRUNPARIS2}
Charm thermal relaxation time constant obtained for 
$\alpha_{s}(M_{\rm Z})=$ 0.118,
$m_{\rm c}(M_{\rm Z})= 0.7\pm7\%$~GeV.}
\end{figure}

\section{Statistical hadronization}\label{stathad}
Hadron production is well 
described in a large range of yields by the mechanism 
of statistical hadronization~\cite{Tor03}. This 
approach works in high energy nuclear 
collisions far better than has been anticipated by Fermi~\cite{Fer53},
the inventor of this method. An important result of 
statistical hadronization is that 
within a particle `family',  particle yields with same valance quark
content are in relation to each other 
thermally  equilibrated, e.g., the relative yield of 
$\Delta(1230)(qqq)$ and $N(qqq)$ or $K^*(\bar s q)$ and $K(\bar s q)$
are completely controlled
by the particle masses $m_i$ ,  statistical weights (degeneracy) $g_i$ and the 
hadronization temperature $T$. In the Boltzmann limit
one has: 
\begin{equation}\label{RRes}
{n^*\over n}= {g^*m^{*\,2}K_2(m^*/T)\over g\,m^{2}K_2(m/T)}.
\end{equation}

Measurement of hadron resonances can thus sensitively test 
the statistical hadronization hypothesis. Since it is possible that the
decay products of resonances rescatter, in principle not all 
resonances can be reconstructed by the invariant mass method. Thus,
one expects that the experimental result must be  below the ratio \req{RRes}
expected for a given temperature. This will be
the case for our study of $K^*/K$ and $\Lambda(1520)/\Lambda$ for which the
data is available. In fact, in our study of particle yields 
in section~\ref{SPSRHIC},
the experimental ratio~\cite{Adl02pc,Fac02iz,Zha03}: 
$(K^*+\overline{K^*})/K^-=0.26\pm0.07$.
We find that this result is 
exactly in agreement with  our hadronization
temperature, $T=145$ MeV, see table~\ref{RHIC-130}. 
We also find $\Lambda(1520)/\Lambda=0.053$ which is about 
twice as large as the experimental value~\cite{Mar02xi}: $0.025\pm0.07$. 
This reduced experimental
yield confirms that d-wave resonance $\Lambda(1520)$ 
is particularly fragile in matter~\cite{Raf01hp}.

The yields of particles are aside of temperature also controlled by 
their fugacity $\Upsilon_i\equiv e^{ \sigma_i  /T}$, where $ \sigma_i$
 is particle `i' chemical potential. Since for each related  
particle and antiparticle  pair we need two chemical potentials, it 
has become convenient to choose parameters such that we can control 
the difference and sum of these separately. For example for nucleons
and antinucleons $N,\overline{N}$ the
two chemical factors are  chosen as: 
\begin{equation}
\sigma_{N}\equiv \mu_b +T\ln\gamma_N ,\qquad
\sigma_{\overline{N}}\equiv -\mu_b +T\ln\gamma_N,
\end{equation}
\begin{equation}
\Upsilon_N=\gamma_N e^{ \mu_b /T}, \qquad\qquad
\Upsilon_{\overline{N}}=\gamma_N  e^{- \mu_b /T}.
\end{equation}

The role of the two factors can be understood
considering at the first  law of thermodynamics: 
\begin{eqnarray*}
dE+P\,dV-T\,dS&=&\sigma_N\,dN+
      \sigma_{\overline{N}}\,d\overline{N},\\
&=&
 \mu_b (dN-d\overline{N})+ T\ln \gamma_N (dN+d\overline{N}).
\end{eqnarray*}
The   (baryo)chemical potential 
 $\mu_b$,  controls the baryon number, arising from the particle difference. 
$\gamma$, the phase space occupancy,   regulates the number of nucleon--antinucleon pairs present. At the quark level (combining the light
quarks $u,d$ in one $q$, we have the situation shown in table~\ref{parameters}.
\begin{table}[t]
\caption{\label{parameters}Four quarks: $s,\ \overline{s},\ q,\ \overline{q} \to$
 four chemical parameters.} 
\begin{indented}\item[]
\begin{tabular}{ll|l}
\br $\gamma_{i}$&   controls overall abundance & Absolute   chemical\\
&of quark  ($i=q,s$)  pairs & equilibrium\\
\mr
 $\lambda_{i}$&   controls difference between & Relative   chemical\\
&strange and non-strange quarks ($i=q,s$) & equilibrium\\
\br
\end{tabular}
\end{indented}
\end{table}

There is considerable difference how the two types
 of chemical factors influence particle 
yield equilibration. This is best understood 
considering  strangeness in the hadronic gas phase, 
the two principal chemical processes are seen in \rf{exchange}.
The  redistribution of strangeness among 
(in this example) $\Lambda,\,\pi$ and $N,$ K
constitutes approach to the relative  chemical equilibrium 
of these species. The 
 production processes, on right in \rf{exchange}, 
are responsible for absolute   
chemical equilibrium of strangeness. Achievement of the 
absolute equilibrium, $\gamma\to 1$, require  more rarely 
occurring truly inelastic collisions with creation of new particle
pairs. These processes are absent in usual 
nonrelativistic chemical environments.

\begin{figure}[h]
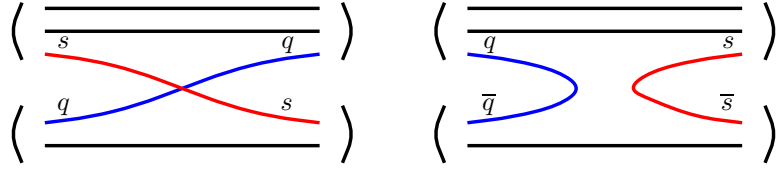
 
\begin{center}
\hspace*{1.9cm}\psfig{width=4.5cm,figure=\pathnow exchange.eps}
\hspace*{1cm}\psfig{width=4.5cm,figure=\pathnow produc.eps}
\end{center}
\vspace{-2.2cm}
\hspace*{3cm}${ s}$ \hspace*{2.7cm}${ q}$
\hspace*{2.4cm}${q}$\hspace*{3.0cm}${ s}$\\[0.4cm]
\hspace*{3cm}${ q}$\hspace*{2.8cm}${ s}$
\hspace*{2.4cm}${ \overline{ q}}$\hspace*{3.0cm}${ \overline{ s}}$
\vspace{0.5cm}
\caption{\label{exchange}
Typical strangeness exchange (left) and production (right)
reactions in the hadronic gas phase.}
\end{figure}

When statistical hadronization of, e.g., a deconfined state
of QGP occurs,  the hadron  yields 
are following closely the magnitude of the accessible phase space
characterized by these chemical factors (and temperature). An important
feature is the  apparent under or
over population of the resulting phase space.  Namely, even 
if there is the pair abundance equilibrium in the primary QGP
phase, in the secondary phase (here hadronic gas) 
phase space has different  size in general. 
Absolute chemical equilibrium in the
secondary phase requires  a period of pair production 
accompanying hadronization. A long lasting QGP--HG  mixed phase is 
today ruled out by experiment, and we should expect
that the phase space occupancy chemical factors $\gamma_i$ 
will be discontinuous at the phase transformation, and in general 
in the observed final state $\gamma_i\ne 1$. 

There are many different hadrons, and in principle, we should
assign to each a chemical potential and than look for 
chemical reactions which relate these chemical potentials,
e.g., on left in \rf{exchange}, we infer
$\mu_\Lambda+\mu_\pi=\mu_N+\mu_K$. 
However, more direct way to accomplish
the same objective consists in  characterizing 
each particle by the valance quark content~\cite{Koc83}, 
forming a product of chemical factors,  e.g.,  for  $p(uud)$,
\[
\Upsilon_{p(uud)}=\gamma_u^2\gamma_d\ \lambda_{u}^2\lambda_{d},\qquad\qquad
\Upsilon_{\bar p(\bar u\bar u\bar d)}=\gamma_u^2\gamma_d\ \lambda_{u}^{-2}\lambda_{d}^{-1},
\] 
note here that:  
\[
\lambda_{i}=e^{\mu_{i}/T},\   \qquad
\mu_q=\frac12 (\mu_u+\mu_d),\quad \lambda_q^2=\lambda_u\lambda_d\quad
\lambda_b=\lambda_q^3.
\]
This implies relations between quark based $\mu_i, i=u,d,s$
 and hadron based $\mu_i, i=b,S$ chemical potentials:
\[
\mu_b=3\mu_q\qquad
\mu_s=\frac 1 3 \mu_b-\mu_{\rm S},
\qquad \lambda_s={\lambda_q\over\lambda_{\rm S}},
\]
An important anomaly is the historically  negative 
S-strangeness in $s$-hadrons. e.g.:
\[
\Upsilon_\Lambda=\gamma_u\gamma_d\gamma_s\,e^{(\mu_u+\mu_d+\mu_s)/T}
           =\gamma_u\gamma_d\gamma_s e^{(\mu_b-\mu_S)/T},
\]
\[
\Upsilon_{\overline\Lambda}=\gamma_u\gamma_d\gamma_s\,e^{(-\mu_u-\mu_d-\mu_s)/T}
   =\gamma_u\gamma_d\gamma_s\,e^{(-\mu_b+\mu_S)/T}.
\]

The phase space density is:
\begin{equation}
 {{d^6N_i}\over{d^3pd^3x}}=g_i{ \Upsilon_i  
\over (2\pi)^3}e^{-E_i/T},
\end{equation}
\begin{equation}
 {{d^6N_i^{\rm F/B}}\over{d^3pd^3x}}= {g_i\over (2\pi)^3}
{1\over  \Upsilon_i^{-1} e^{E_i/T}\pm 1},\quad
 \Upsilon_i^{\small\rm bosons} \le e^{m_i/T}.
\end{equation}
and thus the $4\pi$ particle yield is 
 proportional to the  phase space integral, for example for $\pi,\ N$ and $\overline N$:
\[
\ds \frac{ N_\pi }{V} = C
g_\pi\!\!\int\!\!\frac{d^3p}{(2\pi)^3}\frac{1}{\ga_q^{ -2}e^{\sqrt{m_\pi^2+p^2}/T}-1}\,,
\quad \ga_q^2<e^{m_\pi/T}\simeq (1.6)^2,
\ds  
\]
\[
\fl \hspace*{1cm}
\ds \frac{ N }{V} =C 
g_N\!\!\int\!\!\frac{d^3p}{(2\pi)^3}\frac{1}{1+\ga_q^{ -3}\la_q^{ -3 }
e^{E/T}},
\quad \frac{ \overline {N}}{V} =
Cg_N\!\!\int\!\!\frac{d^3p}{(2\pi)^3}\frac{1}{1+\ga_q^{ -3 }\la_q^{ +3 }
e^{E/T}}\ds. 
\]
One can show that there is no influence of matter flow dynamics
on this complete `$4\pi$' particle yield. 

When a small region of rapidity is considered, but the yield 
of particles is practically constant 
as function of rapidity, it is possible to imagine that the 
yield arise from a series of fireballs placed at 
different rapidities, and thus in this 
limit we also can proceed as if we had
a full phase space integral. However, the proportionality constant  $C$ 
is, in this case in particular, but also more generally 
for a dynamically evolving system,
not the system  volume. We recall that though the individual phase
space integrals are easily evaluated, to get the full yield 
one has to be sure to include all the hadronic 
particle decays feeding into the  yield considered, e.g., the decay 
 $K^*\to K+\pi$ feeds into $K$ and $\pi$ yields. 
This actually constitutes a book keeping challenge in study of particle 
multiplicities, since decays are contributing at the 50\% level to 
practically all particle yields,  sometimes the decay contribution
can be dominant, as is generally the case for the  pion yield,
yet each resonance contributes relatively little in the final count, it is
the large number of resonances that matters. 
 
It is often more appropriate to study ratios of particle yields as these
can be chosen such that certain physical features can be isolated. For
example, just the two ratios  
\[
R_\Lambda = 
\frac{\overline{\Lambda}+\overline{\Sigma}^0+\overline{\Sigma}^*+\cdots }{
{\Lambda}+{\Sigma}^0+{\Sigma}^*+\cdots}=\frac{\bar s\bar q\bar q}{sqq}=
\lambda_{\rm s}^{-2} \lambda_{\rm q}^{-4} =
e^{2\mu_{\rm S}/T}e^{-2\mu_b/T}\, ,
\]
\[
R_\Xi =  
\frac{\overline{\Xi^-}+\overline{\Xi^*}+\cdots }{{\Xi^-}+{\Xi^*}+\cdots}=
\frac{\bar s\bar s\bar q}{ssq}=
 \lambda_{\rm s}^{-4} \lambda_{\rm q}^{-2}=e^{4\mu_{\rm S}/T}e^{-2\mu_b/T} \, ,
\]  
lead to a very good estimate of the baryochemical potential and strange
chemical potential~\cite{Raf91a}, and thus to predictions of other particle 
ratios. The   sensitivity to phase space occupancy 
factors $\gamma_i$  derives from comparison of hadron yields
with differing $q,s$ quark content, e.g.:
\[ 
\frac{\Xi^-(dss)}{\Lambda(dds)}\propto 
\frac{\gamma_d\gamma_s^2}{\gamma_d^2\gamma_s}\
\frac{g_\Xi\lambda_d\lambda_s^2}{g_\Lambda\lambda_d^2\lambda_s}\,,\qquad
\frac{\overline\Xi^-(\bar d\bar s\bar s)}{\overline\Lambda(\bar d\bar d\bar s)}
\propto 
\frac{\gamma_d\gamma_s^2}{\gamma_d^2\gamma_s}\
\frac{g_\Xi\lambda_d^{-1}\lambda_s^{-2}}{g_\Lambda\lambda_d^{-2}\lambda_s^{-1}}\,.
\] 
Note that  $\gamma_q^2\equiv \gamma_u\gamma_d$ and $\gamma_u\simeq\gamma_d$.

An interesting application   arises when we consider product of the above
ratios, as than the result depends on $\gamma_s/\gamma_d$ and temperature. 
In \rf{JRPLGSGQT3}, we see how this product of ratios  shows for the RHIC$_{130}$ 
results that the equilibrium value $\gamma_s/\gamma_q=1$ 
(where we set $\gamma_d\simeq\gamma_u\simeq\gamma_q$) is not  compatible with the
experimental results. Recall that the phase space occupancies 
we consider here are those arising from hadron yields and thus are 
applicable to the hadron yields, the excess above chemical equilibrium 
does not imply that the underlying state of, e.g., deconfined quark--gluon plasma is 
not chemically equilibrated. 

\begin{figure}[tbh]
\hspace*{2.5cm}\psfig{height=9cm,clip=1,
figure=\pathnow 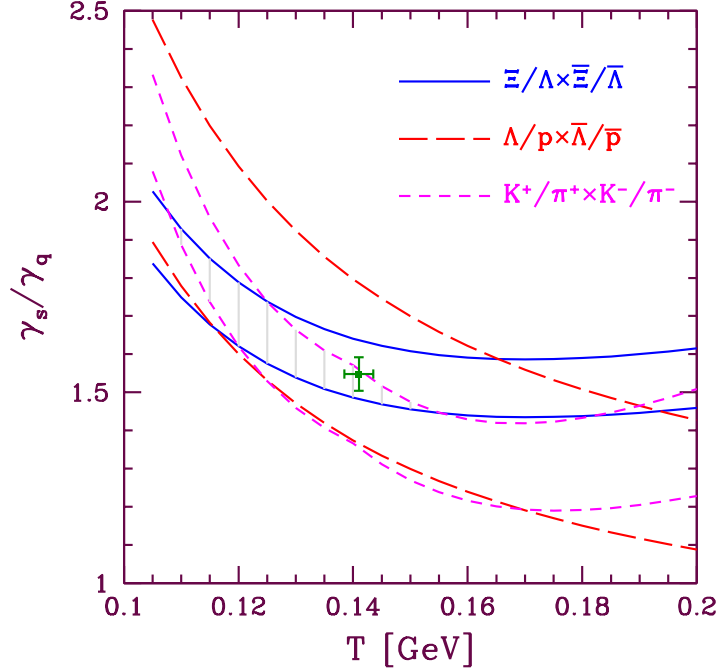}
\vspace*{-0.3cm}
\caption{\label{JRPLGSGQT3}
Evaluation of results of RHIC$_{130}$ experimental particle ratio results 
showing that $\gamma_s/\gamma_q>1$~\cite{Raf02ga}. The cross is result of 
a global fit to all data, the bands reflect on the experimental uncertainty. 
Hatched area is the compatibility region for 
the three particle ratio products considered.  
}
\end{figure}

It is at first hard to believe that the overpopulation of the strange quark 
phase space, as shown  in \rf{JRPLGSGQT3}, should be 50\% grater than that of 
light quark phase space. However, this should have been expected! As the 
kinetic theory of strangeness production presented in  section \ref{sprod}
shows, the lower mass quarks approach equilibrium yield faster. Thus, while
practically all QGP produced strange quark pairs escape to be observed,
light quark pair can be more easily reequilibrated in their abundance.
This argument assumes that the hadronization is not entirely sudden. 

In the limit of very
rapid hadronization, another mechanism favors abundance of 
strange quark pairs over light quark pairs in that there is a
relatively small upper limit on $\gamma_q$. 
As there is no time to expand the volume, hadron formation has 
to absorb the high
entropy  content of QGP which originates in broken color bonds.
As is seen in \rf{JRABSSNE}, the maximum  entropy 
density $S/V$ occurs for  an oversaturated  pion gas, 
$\gamma_q\simeq e^{m_\pi/2T}\simeq 1.6$. Just below 
this Bose condensation condition, the entropy density is
twice as large  as that of
chemically equilibrated gas. The entropy content of a 
non-equilibrium Bose gas has to be evaluated 
recalling 
\begin{equation}
{S}_\pi=
 \int\!\frac{d^3\!p\, d^3\!x}{(2\pi\hbar)^3}\,
   \left[(1+f_\pi)\ln(1+f_\pi)-   f_\pi\ln f_\pi\right]\,,
\end{equation}
\begin{equation}
f_\pi(E)=\frac{1}{\gamma_{q}^{-2}e^{E_\pi/T}-1}
\,,\quad 
E_\pi=\sqrt{m_\pi^2+p^2}.
\end{equation}
We note that
a similar constrain arising from kaon condensation 
limit yields the less restrictive condition:
\begin{equation}\label{Kcond}
{\gamma_s\over\gamma_q}\left(\lambda_s\over\lambda_q\right)^{\!\!\pm 1}
<e^{(m_{\rm K}-m_\pi)/ T}\simeq 11\,.
\end{equation}
We return to this interesting condition in section~\ref{lhc}.

\begin{figure}[h]
\vspace*{1.9cm}\hspace*{1.cm}
\epsfig{width=10cm,height=7cm,figure=\pathnow 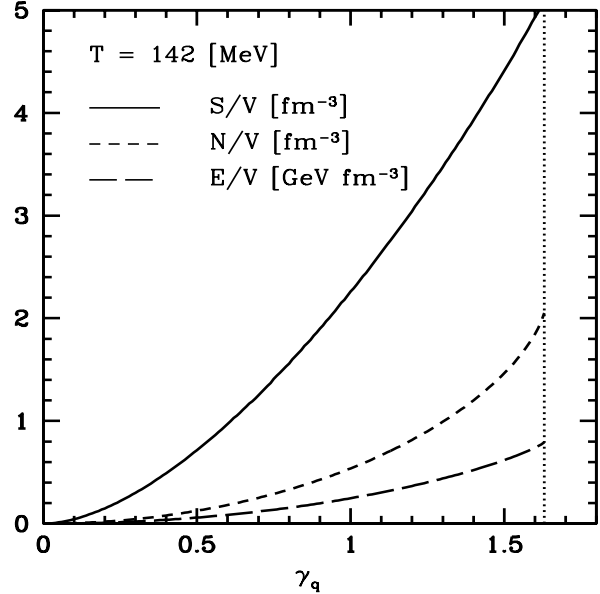}
\vskip -.8cm
\caption{\label{JRABSSNE}
Entropy density $S/V$ along with particle density $N/V$ and energy density 
$E/V$ for a non-equilibrium pion gas at $T=142$\,MeV. 
}
\end{figure}

\section{Hadronization at  SPS and RHIC}\label{SPSRHIC}
Given yields of (strange) hadrons, we can apply the principles described 
above to study the hadronization conditions, and we would like to do this 
here as function of energy. Experimental results were
available to consider RHIC Au--Au collisions obtained at 
$\sqrt{s_{NN}^{CM}}={130}$ GeV  and SPS Pb--Pb reactions at 
$\sqrt{s_{NN}^{CM}}=8.75, 12.25,17.2$~GeV  
(projectile energy 40$A$ GeV, 80$A$ GeV, 158$A$ GeV), 
as well as  S--Pb/W 200$A$ GeV reactions occurring at 
$\sqrt{s_{NN}^{CM}}=19.2$ GeV. The procedure we adopt is
to fit statistical parameters for $4\pi$ particle 
multiplicity results. For SPS, the resulting $\chi^2/$dof is shown
in \rf{PLCHIEB}. The solid squares are obtained allowing 5 parameters,
i.e., full chemical nonequilibrium,  the open triangles assuming that complete
chemical equilibrium prevails. Open square results were obtained 
assuming that light quarks are in 
chemical equilibrium.  Except at  40$A$ GeV, we find  in our approach 
involving chemical nonequilibrium yields and only statistical 
errors, statistically very significant fits. 

We use here SPS NA49 results~\cite{Gaz03}, which include 
$\pi^\pm,\ K,\ \overline{K},\ \Lambda,\ \overline{\Lambda},\ \phi$ at
40, 80, 160$A$ GeV and  at top SPS energy, we find that the experimental
results for  $\Xi,\ \overline\Xi,\ \Omega,\ \overline\Omega$ are reproduced
nearly exactly even though  for consistency we have not fitted these. 
Since this is a different data sample and we do not  employ presently the 
central rapidity WA97 results, our fit result for some parameters 
will be slightly different, e.g., though we still report $\gamma_q^{\rm HG}>1$,  
the value we find is  smaller than found using the complete data sample of
NA49 and WA97. 

\begin{figure}[t]
\hspace*{2.5cm}\psfig{width=10cm,
figure=\pathnow  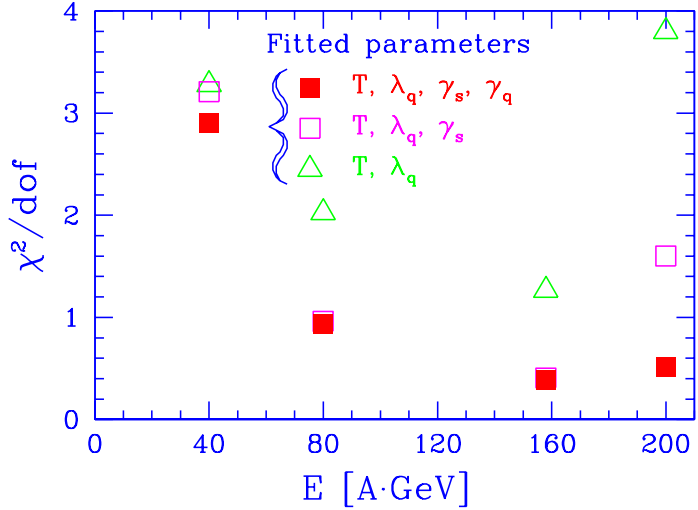}
\caption{\label{PLCHIEB}
Quality of statistical hadronization fit to hadron yields: $\chi^2$/dof
for the 4 SPS reaction energies.
}
\end{figure}
\begin{table}[t]
\caption{\label{RHIC-130}
Results of fit of RHIC$_{130}$ particle yields. Top line: statistical significance,
following sections: fitted statistical parameters, hadro-chemical
potentials, physical properties of the fireball. Stars indicate that a value is
consequence of a constraint, and was not fitted. 
}
\begin{indented}\item[]
\begin{tabular}{l|c|c}
\br
RHIC$_{130}$                  & non-equilibrium     & equilibrium\\
\mr    
$\chi^2/$dof                     &27/($21-3$)        & 230/($21-2$) \\
\mr    
$T${\vphantom{$\frac AB$}}      & 144.6 $\pm$ 1.3    & 169.1 $\pm$ 2.3   \\
$\lambda_{ q}$                  & 1.069 $\pm$ 0.008 & 1.067 $\pm$ 0.008\\
$\lambda_{ s}$                  &  1.0198$^*$        &  1.0167$^*$  \\
$\gamma_{ q}^{\rm HG}$          & 1.62$^*$           &  1$^*$            \\
$\gamma_{ s}^{\rm HG}/\gamma_{ q}^{\rm HG}$  &  1.53 $\pm$ 0.03 & 1$^*$ \\
\mr{\vphantom{$\frac AB$}}
$\mu_b$ [MeV]                   &29.1                & 32.8    \\
\ $\mu_{\rm S}$ [MeV]           &6.9                 & 8.1      \\
\mr{\vphantom{$\frac AB$}}
 $s/b$                          &  8.8               &   6.2      \\
 $E/b$ [{\small GeV}]    $\!\!$ &32.8                &   31.8    \\
$S/b$                           & 215                &  218     \\
$E/S$ [{\small MeV}]   $\!\!$   & 153                &  146       \\
\br
$W_s $                          & 0.55               &0.49\\
\br
\end{tabular}
\end{indented}
\end{table}

The RHIC fit favors non-equilibrium by such a large
margin that we could not show the result on the same scale in \rf{PLCHIEB}.
The actual values for  $\chi^2$/dof are seen 
in the top line of \rt{RHIC-130}. In obtaining these results it is 
assumed that 40\% of weak decay cascades  $\Xi\to \Lambda$  and $\Lambda\to p$
are accepted whenever such corrections were not applied to data by the
experimental groups. In the chemical nonequilibrium fit,
we assume that the maximum value of $\gamma_{ q}=e^{m_\pi/2T}$ is attained,
and thus this value is not fitted.
The value of $\lambda_{ s}$  is obtained from the
requirement that strangeness balances, $\langle s-\bar s\rangle=0$.
The other statistical parameter acquire values seen in  \rt{RHIC-130}.
We note that the hadronization temperature is greatly reduced in non-equilibrium 
approach compared to the value arrived at forcing chemical 
equilibrium. This is  made possible by $\gamma_i>1$, in this case 
the high particle yields seen at RHIC can be 
arrived at a lower value of hadronization temperature. We also present
the hadron gas parameters $\mu_b, \mu_{\rm S}$ derived from 
$\lambda_q, \lambda_s$ which can be used
in back of envelope cross checks of expected particle ratios.
These results constitute a slight update of our standard RHIC fit~\cite{Raf02ga},
where the actual particle multiplicities are also given.

Given the fitted set of statistical 
parameters ($T,\mu_i,\gamma_i$) which 
characterize the phase space, we can evaluate, 
up to a common normalization constant, 
the  strangeness $\langle s\rangle\simeq \langle \bar s\rangle $, 
energy $E$,  baryon number $b$, entropy $S$ contained by all produced 
hadronic particles, as shown in \rt{RHIC-130}. At the very bottom,
we also present the Wr\'oblewski ratio~\cite{Wro85}, 
\begin{equation}\label{Wro}
W_s={2\langle s\bar s\rangle\over \langle d\bar d+u\bar u\rangle},
\end{equation}
of newly produced strange 
quark pairs to all quark pairs. Strangeness production, while enhanced
is not yet as abundantly produced as light flavors.
We consider this ratio as function of $\sqrt{s}$ in \rf{WSnoe}.
The triangles  for elementary $p$--$p$ reactions are consistent with
`color string snapping'  process which yields strangeness pairs with
about  22\% abundance  of the light quark pairs, and 
nearly independent of $\sqrt{s}$.  In heavy ion collisions (circles), 
a much greater and $\sqrt{s}$ dependent result is seen. This
indicates that a new strangeness production 
process arises, which could well be the thermal production
of strangeness in QGP. We  also see, in the background, the results 
of a similar analysis 
obtained within the chemical semi-equilibrium approach~\cite{Bec01a}. 

\begin{figure}[ht]
\hspace*{-0cm}
\hspace*{2.7cm}\psfig{width=8.0cm,clip=,figure=\pathnow 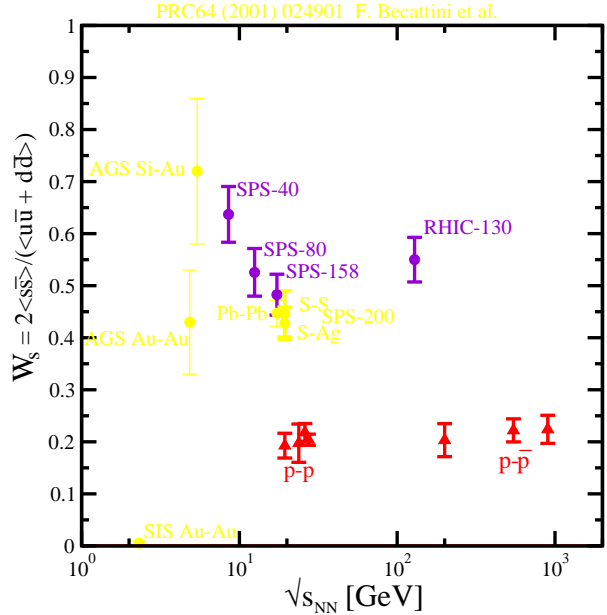}
\vskip -0.5cm
\caption{\label{WSnoe}
Wr\'oblewski ratio~\protect\cite{Wro85} :
only newly made $s$- and $q$-pairs are counted in
comparing  strange  to light quark pair production.
}
\end{figure}

\begin{figure}[t]
\hspace*{3.0cm}\psfig{width=8cm,figure=\pathnow 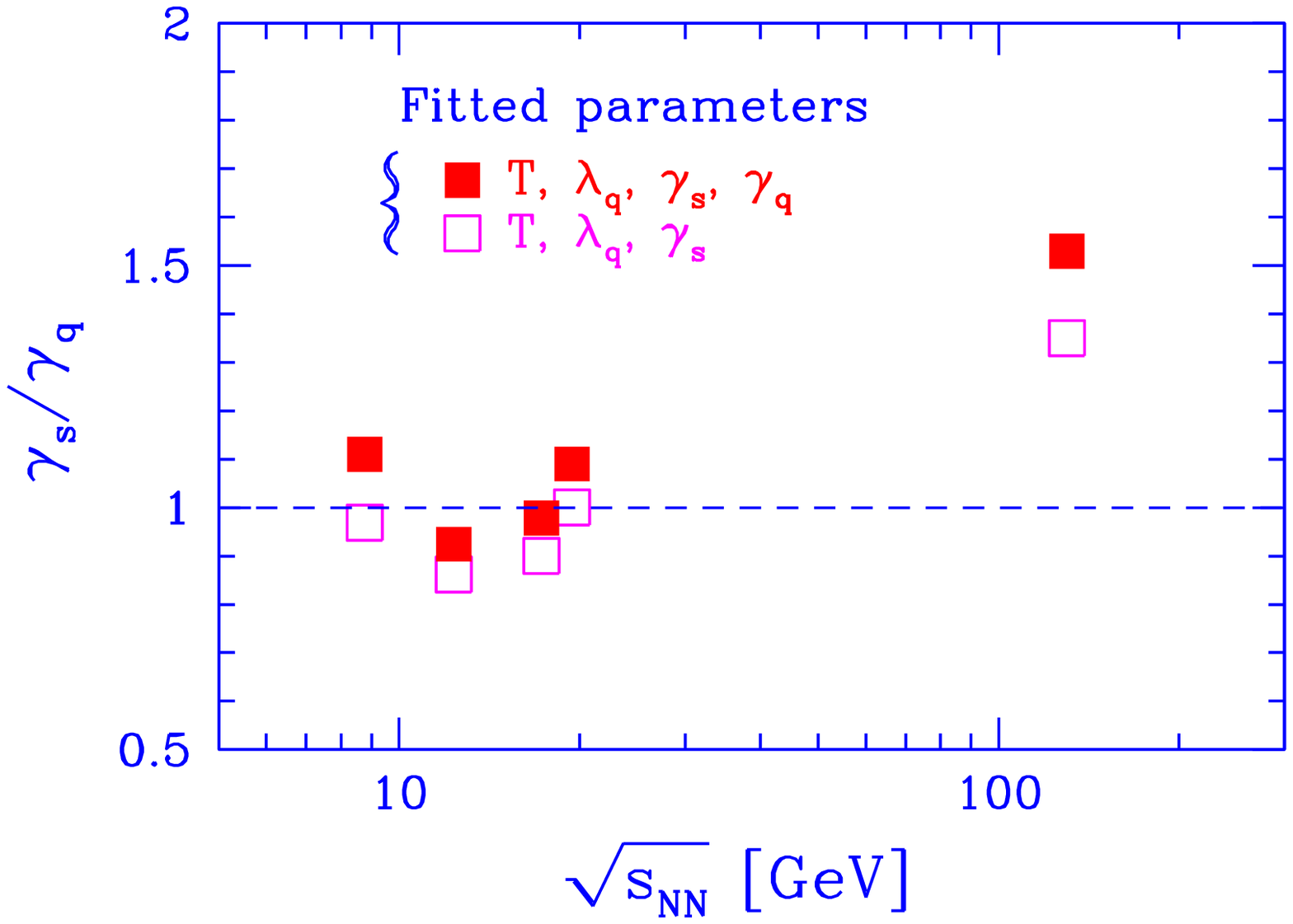}\\
\hspace*{3.0cm}\psfig{width=8cm,figure=\pathnow 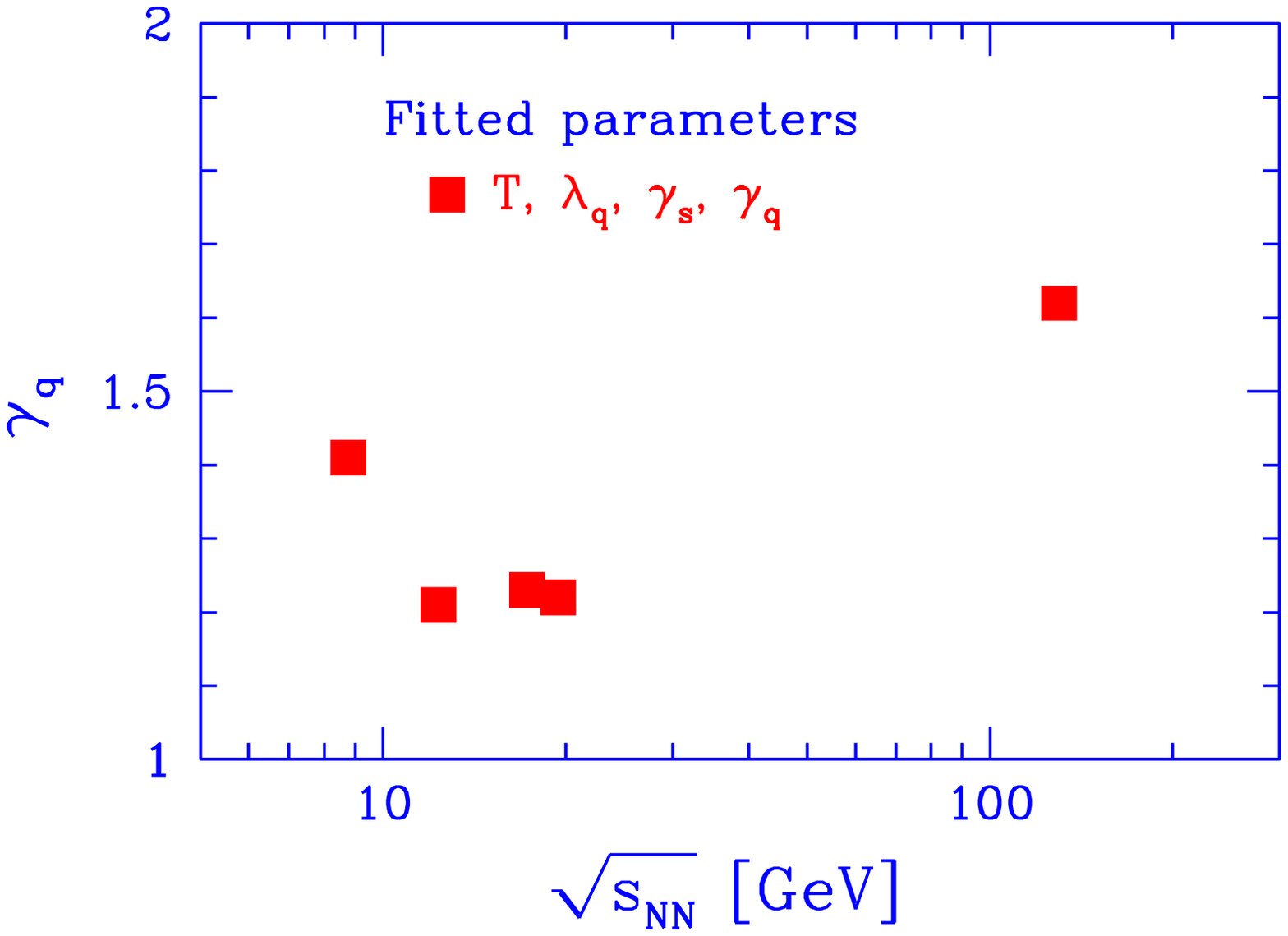}
\vskip -0.5cm
\caption{\label{PLGAQS}
Fits of, top, $\gamma_s/\gamma_q$ and, bottom, $\gamma_q$
chemical nonequilibrium parameters as function of collision 
energy. Open squares results for $\gamma_q=1$, in which case the 
left hand side gives $\gamma_s$. 
}
\end{figure}

The nonequilibrium parameters $\gamma_s/\gamma_q,\ \gamma_q$ 
are displayed in \rf{PLGAQS}.
The results for $\gamma_s/\gamma_q$, on top, show that 
only at RHIC  $\gamma_s>\gamma_q$. Open squares show that 
when we fix $\gamma_q=1$ the SPS fits, in general,  yield 
$\gamma_s<1$, as is widely reported.  However, bottom, we
see that $\gamma_q>1$ in all cases, reflecting on the 
 excess in charged hadron multiplicity.
We note that the 40$A$ GeV SPS result (lowest energy point)
 deviates from the behavior 
systematics of the other results. 

Since we know the incoming energy and we can evaluate the amount of 
thermal energy  found in the final state hadrons, we have a good measure of 
the energy stopping  for the different collision systems ---
to be precise in estimate of the energy stopping,
we would have also to account for the 
energy transfered from intrinsic degrees of freedom to 
the collective flow.  The behavior of the ratio of thermal energy 
to collision energy is seen in \rf{PLEBS}. The result is 
in so far surprising as energy
stopping is somewhat greater at RHIC$_{130}$ than at top SPS energy. 
On the other hand, the rise
of the energy stopping power towards lower SPS energies is expected, 
as is the steep rise  for the asymmetric collision system S--Pb/W.

\begin{figure}[t]
\hspace*{3cm}\psfig{width=8cm,figure=\pathnow  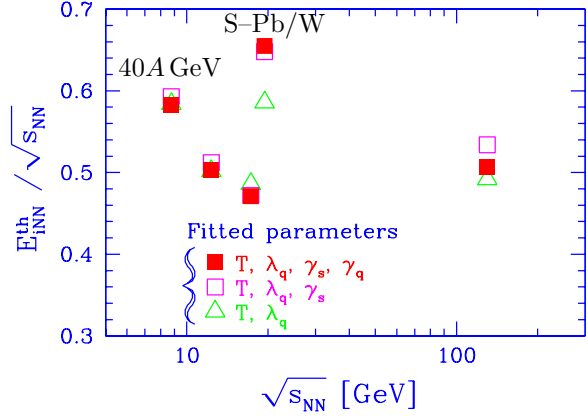}
\hspace*{-.2cm}
\vskip -5.cm
 \hspace*{4.6cm}40$A$\,GeV\vskip-1.0cm
\hspace*{6.cm}S--Pb/W \vskip5.cm
\vskip -0.3cm
\caption{\label{PLEBS}
Fraction of energy stopping at SPS and RHIC: 
results are shown for 40, 80, 158$A$ GeV Pb--Pb,
  200$A$ GeV S--W/Pb reactions and at RHIC for  65+65$A$ GeV Au--Au
interactions. 
}
\end{figure}

\begin{figure}[t]
\hspace*{3.0cm}\psfig{width=7.8cm,figure=\pathnow  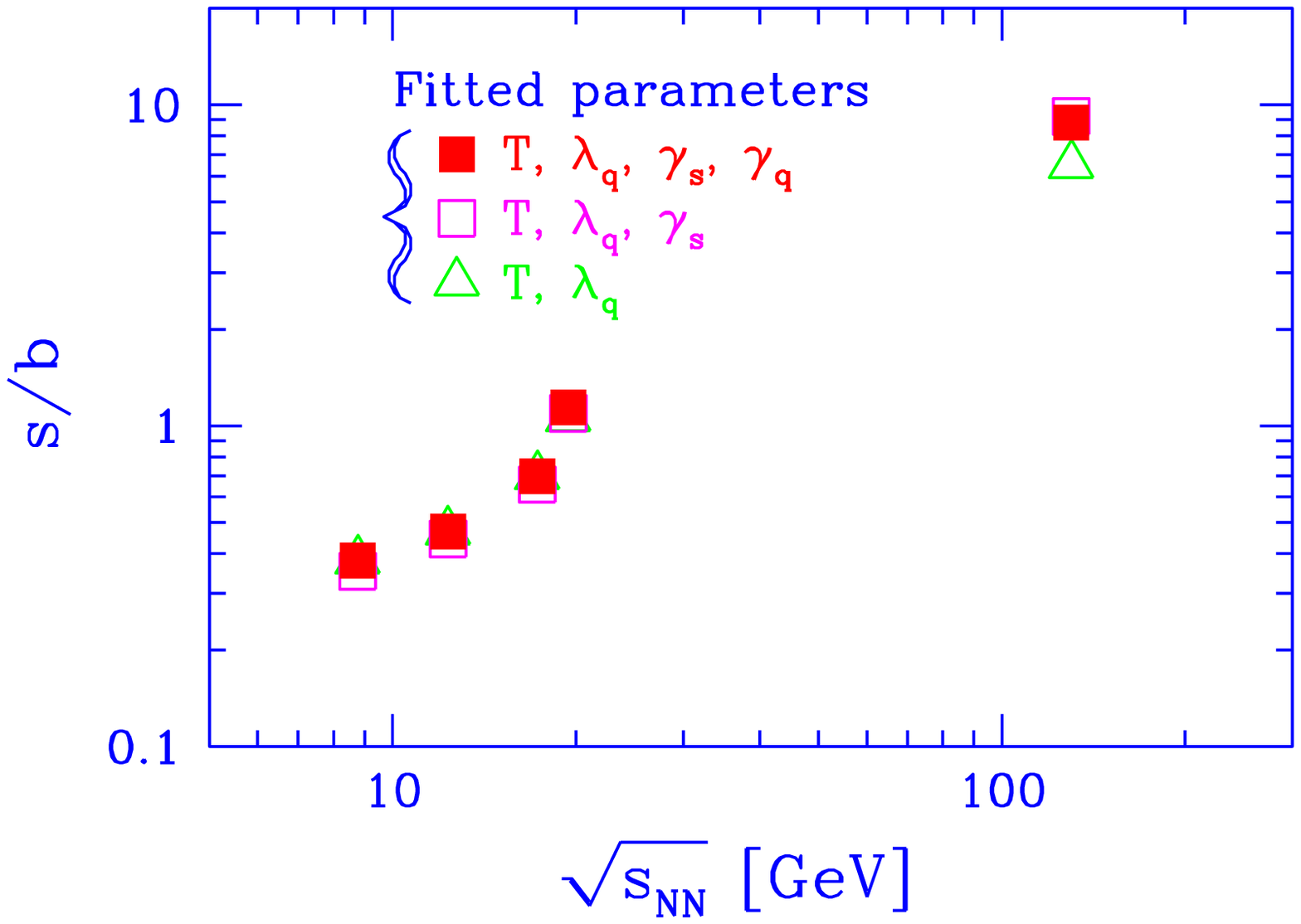}\\
\hspace*{3.0cm}\psfig{width=7.8cm,figure=\pathnow  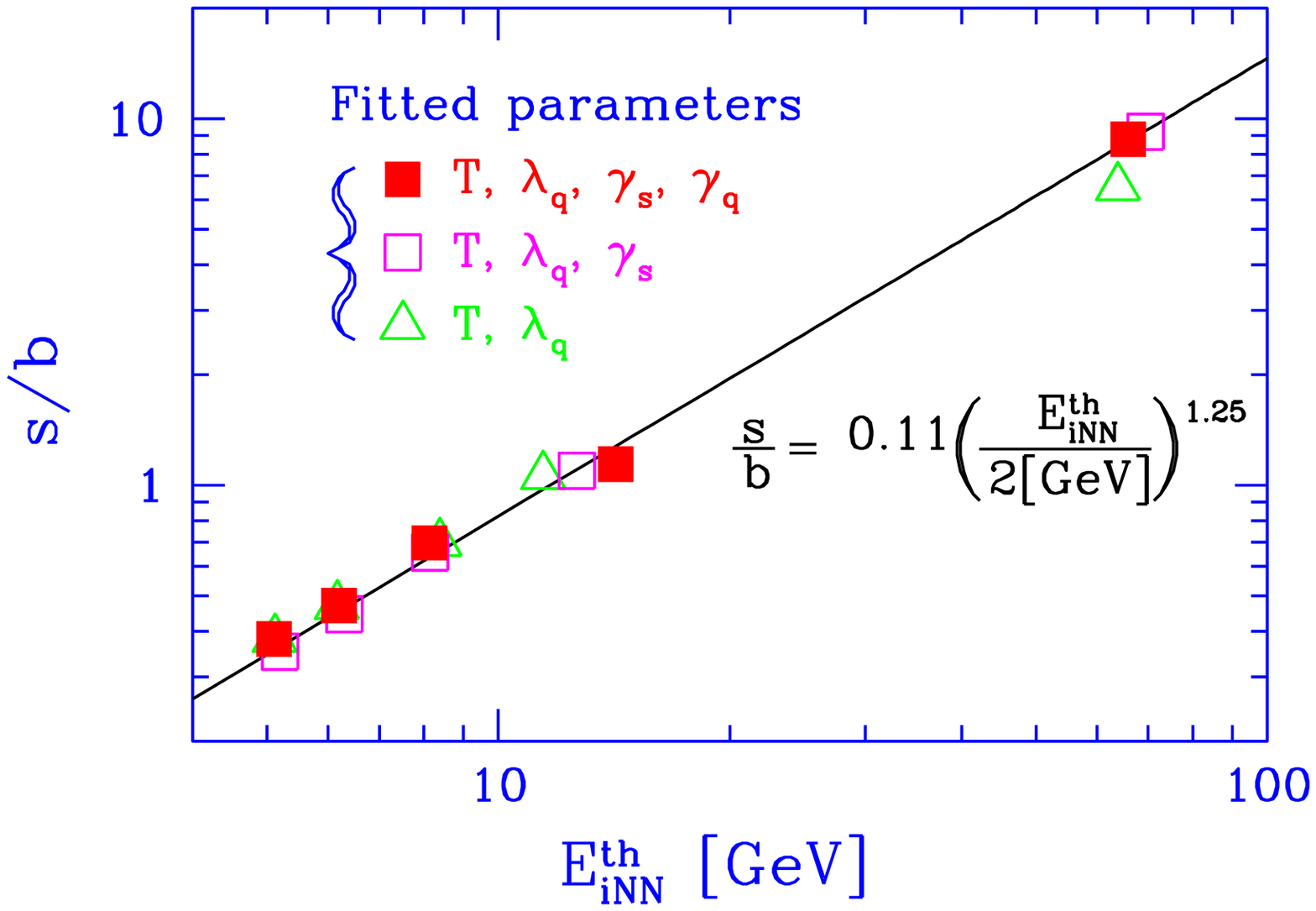}
\vskip -0.4cm
\caption{\label{PLETRBSLOG}
Strangeness per thermal  baryon  participant, top as function of 
 $\sqrt{s_{NN}}$, bottom as function of $E^{\rm th}_{{\rm i}\,NN}$. 
}
\end{figure}

With the relatively great variability of stopping power seen in \rf{PLEBS},
it seems more prudent to use in study of energy dependence of different 
physical properties of the hot fireball not the 
collision energy but the final state 
thermal fireball energy content. This is strongly supported by 
the study of the systematics of strangeness production per baryon, seen
in \rf{PLETRBSLOG}. On top, we show the strangeness excitation function
as usually shown, using $\sqrt{s_{NN}}$ as variable, and normalizing to the
baryon yield. The baryon number is  conserved in hadronization, and
this result is  directly telling us how much strangeness was available prior
to hadronization. We recall that the  true `thermalized' participants are
about 5--10\% fewer than is inferred from geometric 
consideration. On bottom in \rf{PLETRBSLOG},
we present the same result using as variable 
the intrinsic energy content. There is a significant simplification of 
the result and we find
\[
\frac s b =0.11 \left({E^{\rm th}_{{\rm i}\,NN}
          \over 2[\mbox{GeV}]}\right)^{\!5/4}.
\]
Importantly, there is no sign of a deviation  of the systematics 
of strangeness production.  

\begin{figure}[t]
\hspace*{3cm}\psfig{width=8cm,figure=\pathnow  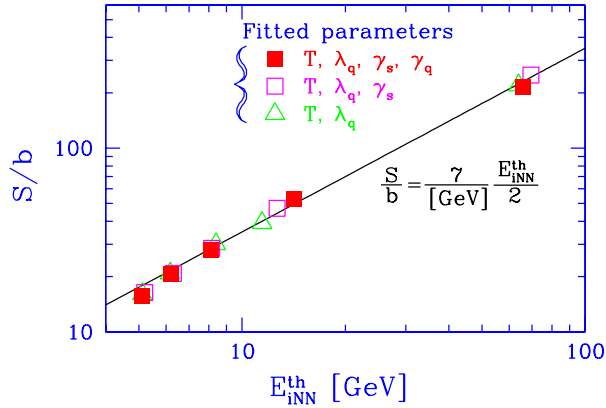}
\vskip -0.4cm
\caption{\label{PLSBEILOG}
Entropy production per thermal participating thermal 
baryon $S/b$ shown as 
function of the intrinsic thermal energy per baryon, 
$E^{\rm th}_{{\rm i}\,NN}$.
}
\end{figure}

Similarly, considering
the specific entropy content in \rf{PLSBEILOG}, we find
that entropy production  per unit of available energy is universal:
\[
\frac{S}{E^{\rm th}_{{\rm i}\,NN}}=\frac{7}{\mbox{[GeV]}}.
\]
The entropy production  mechanism which has this simple
result for a wide range of intrinsic fireball energy is not known, 
but appears to be common for both SPS and RHIC energy range.

Comparing the results seen in~\rf{PLETRBSLOG} and~\rf{PLSBEILOG}, we
note that, as function of the intrinsically available thermal
energy, the  strangeness production rises somewhat faster than entropy,
which expresses the fact that strangeness is more enhanced than 
hadron multiplicity comparing RHIC to SPS.
Within QGP phase, the strangeness to entropy 
ratio $s/S$, shown in \rf{PLETRSS},  characterizes the approach 
to chemical equilibrium of strangeness, as compared to saturation 
of the availability  of quarks and gluons.
 Both quantities may increase slightly in hadronization, 
but are expected to be practically conserved. Since the
rise of $s/S$ from SPS to RHIC is relatively slow, we show the behavior 
as function of both, top 
 $\sqrt{s_{NN}}$, and bottom  of $E^{\rm th}_{{\rm i}\,NN}$. There is 
clear increase of this ratio 
comparing SPS to RHIC, but the magnitude of the 
effect is somewhat dependent on the method of evaluation of chemical 
conditions.

\begin{figure}[t]
\hspace*{3.0cm}\psfig{width=7.8cm,figure=\pathnow  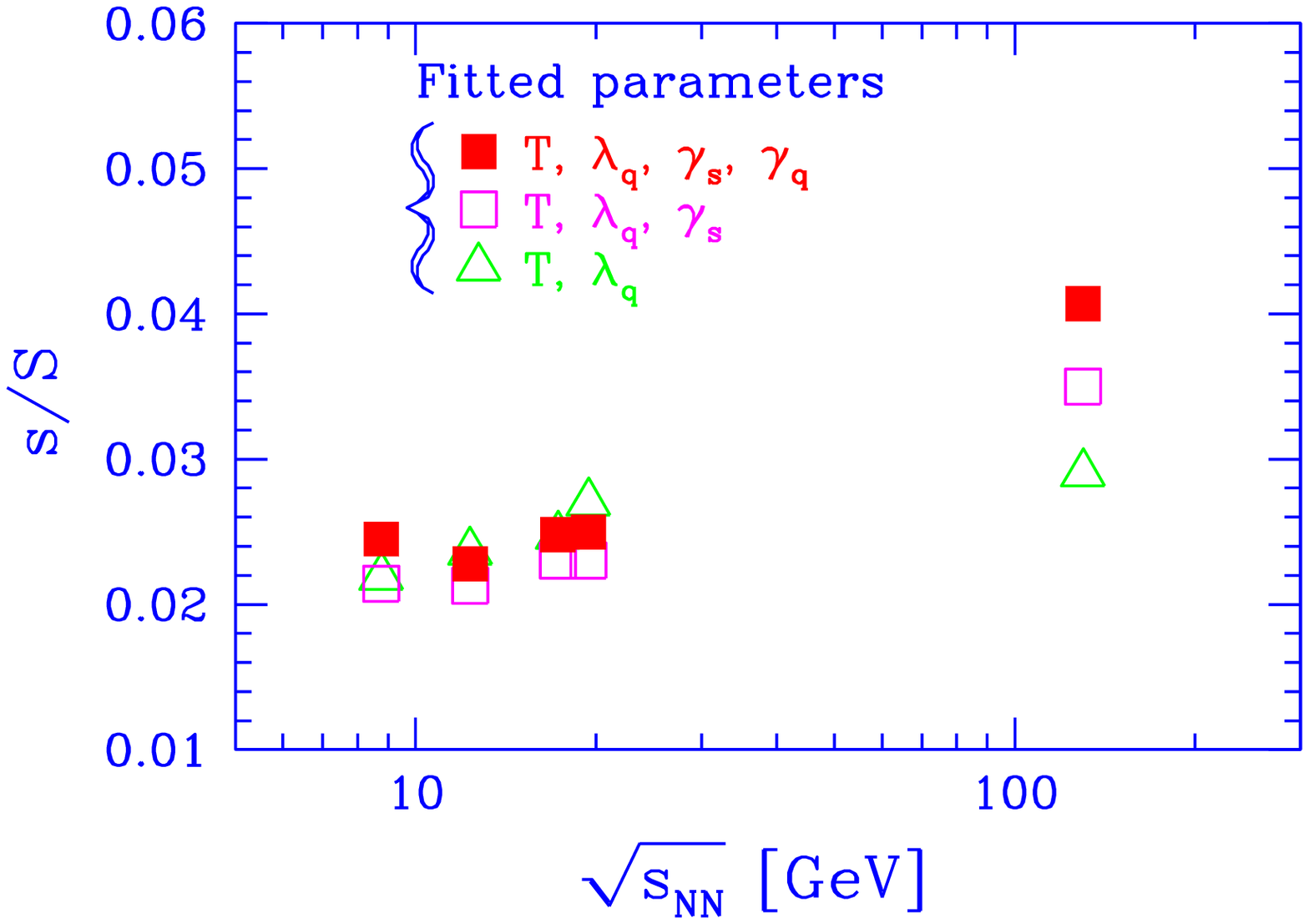}\\
\hspace*{3.0cm}\psfig{width=7.8cm,figure=\pathnow  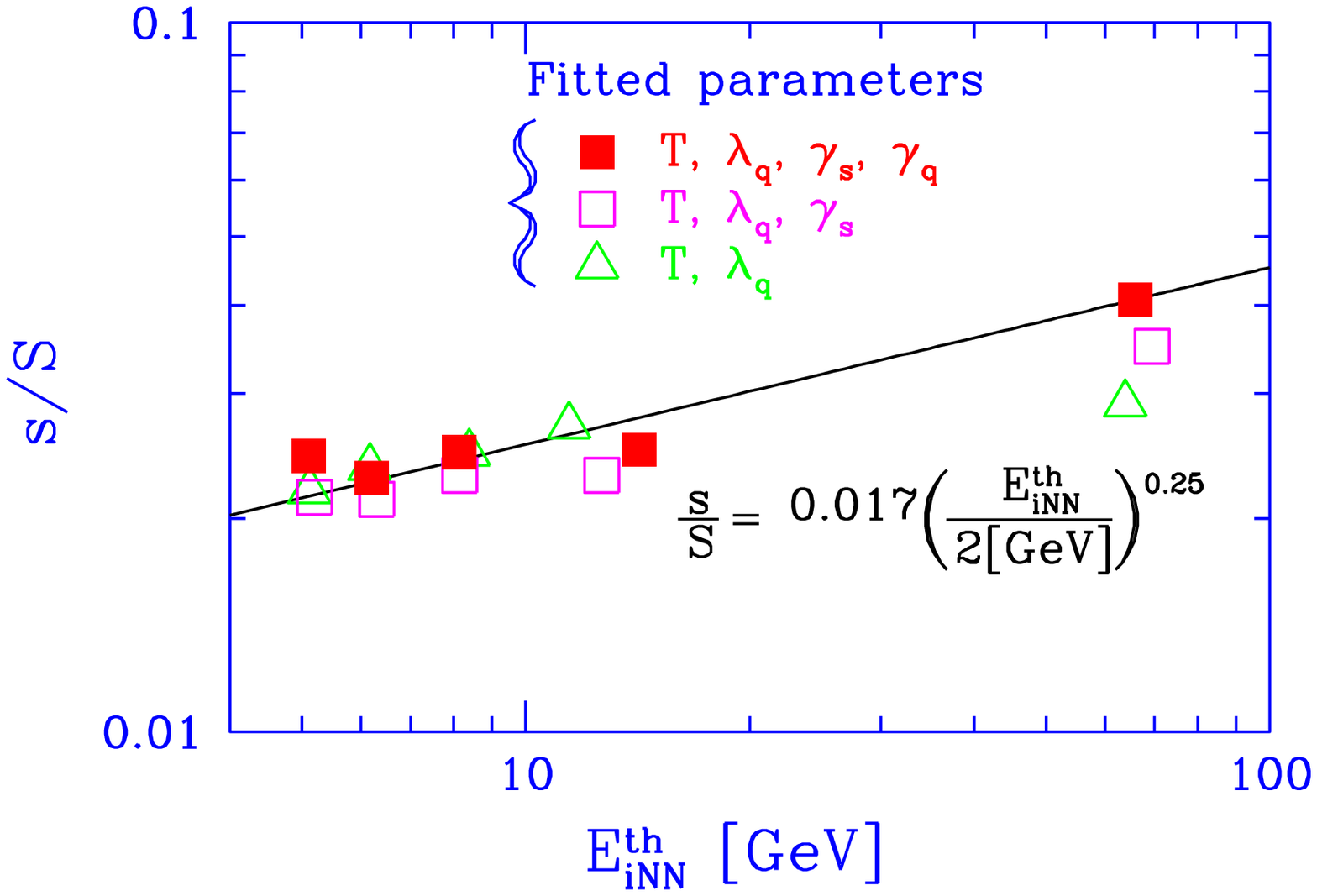}
\vspace*{-0.5cm}
\caption{\label{PLETRSS}
Strangeness per entropy: top as function of
 $\sqrt{s_{NN}}$, bottom  as function of $E^{\rm th}_{{\rm i}\,NN}$.}
\end{figure}
 
There is perhaps an indication, in \rf{PLETRSS}, of
a possible increase of strangeness yield per entropy at the lowest energy,
consistent with the somewhat anomalous rise of $K^+/\pi^+$ ratio 
reported by the NA49 experiment. Dynamically, this may be consequence of 
a longer lived QGP phase, which lacks the strength to undergo a fast
collective flow. Longer lifespan even at lower
initial temperature would help strangeness production. 
More systematic study of strangeness production in this energy
range could thus offer significant insights into the properties of 
the deconfined phase. Ultimately, as the energy is reduced below the 
QGP threshold, there should be a significant cut in strangeness yield.

\section{Onset of quark-gluon plasma formation}\label{HQGP}
The properties of hot QCD matter can be well 
described in terms of a QGP model~\cite{Let03uj},
 which agrees 
with latest lattice results~\cite{Fod02a,Fod02b,Fod02c}.  
Thus, in principle, we understand the behavior of 
the QGP and can  analyze the meaning of the 
relatively low hadronization
temperatures. In \rf{PLTMUBLIQ}, we show with solid squares the SPS
and open square the RHIC hadronization points we obtained, dashed line 
corresponds to the cross over from the deconfined to confined phase.
In the background, we see as triangles hadronization 
analysis for AGS and SIS assuming
equilibrium conditions. 

\begin{figure}[t]
\vspace*{-0.2cm}
\hspace*{3cm}\psfig{width=10cm,clip=,figure=\pathnow 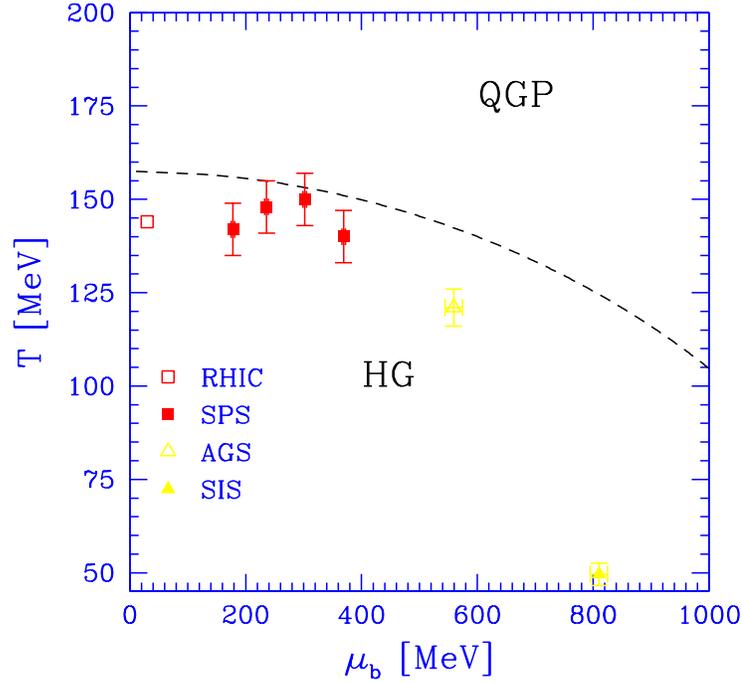}
\vspace*{-0.9cm}
\caption{\label{PLTMUBLIQ}
Hadronization conditions in $T,\mu_b$ plane: RHIC$_{130}$ open square, SPS 
solid squares.
}
\end{figure}
\begin{table}[ht]
\caption{\label{Tmu} The results of nonequilibrium fits at SPS and RHIC: for each
collision system  we present the collision energy in laboratory, reaction
energy $\sqrt{s_{NN}}$, the chemical freeze-out temperature $T$, 
and baryochemical potential $\mu_b$;
these results are shown in \rf{PLTMUBLIQ}.
The bottom line gives the experimental ratio $\mbox{K}/\pi$, where available,
see text for details.}
 \begin{indented}\item[] 
\begin{tabular}{l|cccccc}
\br
                              & Au--Au & S--W & Pb--Pb & Pb--Pb& Pb--Pb& Pb--Pb\\
\mr
$\sqrt{E_{\mbox{lab}}}$\,[GeV]&65+65   & 200   & 158 & 80  & 40  & 30 \\
\mr
$\sqrt{s_{NN}}$\,[GeV]        & 130    & 19.4  &17.2 &12.3  &8.75 & 7.60 \\
\mr
$T$\,[MeV]                    & 144.6  & 143.5 &148.5&150.3&139.5& --\\
\mr
$\mu_b$\,[MeV]                & 30.1   & 165  & 236   & 302   &370  & --\\
\mr
$\mbox{K}/\pi$                & 0.16   & --  & 0.12  &0.11   &0.11 & 0.12\\

\br
\end{tabular}
 \end{indented}
 \end{table}

Our results are tabulated in 
\rt{Tmu}, obtained enforcing strangeness 
conservation (except for S--W) and allowing chemical nonequilibrium. 
As we go to press a fit at 30 GeV is still not possible, with 
particle yields not available. Statistical
errors on the fit results are small and are not shown, in \rf{PLTMUBLIQ} we 
indicate our estimate of the systematic SPS result error based on 5\% individual 
data point error.  The relative changes between the fit results to NA49 40,80,158 GeV 
reactions are most likely significant, since the systematic error is common 
to these results.
Thus it is the 80 GeV SPS hadronization point which is closest to the
equilibrium phase transition between HG and QGP. The RHIC point (open square) 
though at low baryochemical potential could be at lower temperature than
are the 160 GeV SPS and 80 GeV SPS results. The lowest temperature is found for
the S-W point, but this result is obtained within a very different procedure,
using central rapidity results of WA85 and a flow model. 
If at all, the temperature of hadronization seems to rise as the collision
energy falls, until we nearly touch the equilibrium phase transition condition.
From thereon, we record  a different behavior, with hadronization condition at 40 GeV 
moving more inward the hadron domain, with AGS and SPS equilibrium points
deep within hadron gas phase.

To better understand this result, we must remember that the 
fireball is not a piece of deconfined matter 
sitting still. It undergoes a rather 
rapid collective explosive flow. Collective motion of 
color charged quarks and gluons contributes an important 
collective component in the pressure,
as can be seen considering the stress portion of the 
energy-momentum tensor: 
\begin{equation}
T^{ij}=P\delta_{ij}+(P+\varepsilon)\frac{v_iv_j}{1-\vec v^{\,2}}\,.
\end{equation}
The rate of  momentum flow vector  $\vec {\cal P} $ 
at the surface of the fireball
 is obtained from the energy-stress tensor  $T_{kl}$: 
\begin{equation}
\vec {\cal P}\equiv \widehat{\cal T}\cdot \vec n=P \vec n+(P+\varepsilon)
  \frac{\vec v_{\mbox{\scriptsize c}}\, \vec v_{\mbox{\scriptsize c}}\!\cdot\! \vec n}
          {1-\vec v_{\mbox{\scriptsize c}}^{\,2}}\,.
\end{equation}
The pressure  and energy  comprise particle and the  vacuum properties:
 $P=P_{\mbox{\scriptsize p}}-{\cal B}\,,$ $\varepsilon
 =\varepsilon_{\mbox{\scriptsize p}}+{\cal B}\,.$

The condition  $\vec {\cal P}=0$  reads:
\begin{equation}
{\cal B}\vec n=P_{\mbox{\scriptsize p}}\vec n+
      (P_{\mbox{\scriptsize p}}+\varepsilon_{\mbox{\scriptsize p}})
\frac{\vec v_{\mbox{\scriptsize c}} \, \vec v_{\mbox{\scriptsize c}}\!\cdot \!\vec n}
        {1-v_{\mbox{\scriptsize c}}^{2}}\,.
\end{equation}
Multiplying with  $\vec n$ , we find:
\begin{equation}
{\cal B}=P_{\mbox{\scriptsize p}}+
      (P_{\mbox{\scriptsize p}}+\varepsilon_{\mbox{\scriptsize p}})
\frac{\kappa  v_{\mbox{\scriptsize c}}^2}{1-v_{\mbox{\scriptsize c}}^{2}}\,,
\qquad
\kappa=\frac{(\vec v_{\mbox{\scriptsize c}}\cdot \vec n)^2}{v_{\mbox{\scriptsize c}}^2}\,.
\end{equation}
This requires $P_{\mbox{\scriptsize p}}<{\cal B}$: QGP phase pressure 
$P$ must be {\em negative} when the dynamical pressure comprising effect of 
the flow runs its course, ${\cal P}\to 0$. 
 A fireball surface region  which reaches ${\cal P}\to 0$ but 
for which $v\ne 0$, i.e., it continues to flow outward, 
is torn apart in a rapid filamentation instability. 
This situation can {\em only} arise since the quark-gluon 
matter presses again the collective vacuum which is not subject  
to collective dynamics.

Phase boundary between the hadron gas domain and quark-gluon plasma,
 which includes the effect of the  `wind' of flow of QCD matter
is shown in \rf{PLTMUBLIQ3}, for a geometric model with  $\kappa=0.6$. 
With increasing flow velocity the phase boundary 
set at ${\cal P}\to 0$  moves to lower temperatures. This effect is larger
near to $\mu_b=0$ than the effect $\mu_b$ has on change in $T$. 
We can now understand the behavior seen 
in \rf{PLTMUBLIQ}: at RHIC there is the largest supercooling and the flow
velocity is greatest, yielding smallest hadronization temperature. 
As collective flow velocity decreases, the hadronization temperature 
increases, though this effect is somewhat compensated by the influence of the 
the increase in $\mu_b$ which implies a decrease in $T$. The impact of 
the flow velocity is smallest near to  80$A$ collision energy, where 
the hadronization temperature is nearly 
reaching the HG--QGP phase boundary.

\begin{figure}[t]
\vspace*{-1.5cm}
\hspace*{2cm}\psfig{width=11cm,clip=,figure=\pathnow 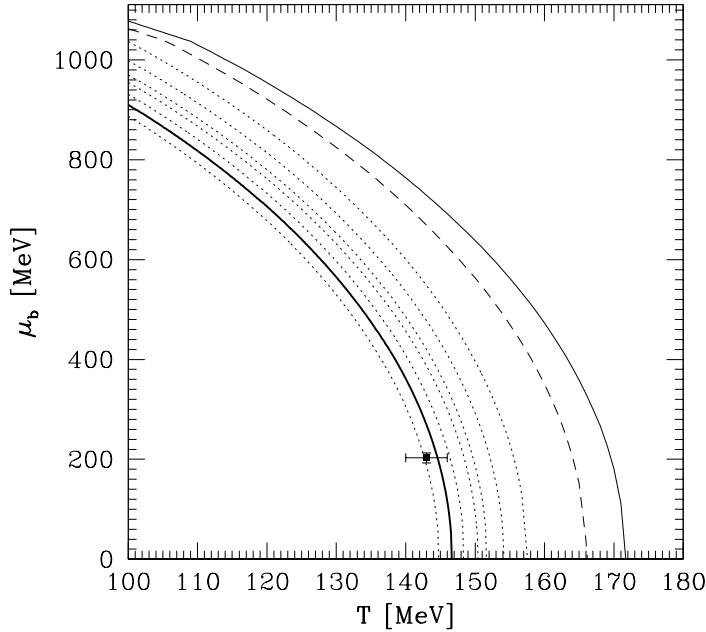}
\vspace*{-1cm}
\caption{\label{PLTMUBLIQ3}
Hadronization boundary in the
$\mu_b,T$ plane: solid (point hadrons)
and dashed (finite volume hadrons) lines 
are estimates of QGP--HG boundary  for a system at rest. 
Dotted lines are for finite size hadrons and include
collective flow velocity for 
$v^2=0,1/10, 1/6, 1/5, 1/4, 1/3$\,.
 Thick solid line: breakup with 
$v=0.54$, $\kappa=0.6$
\protect\cite{Raf00}. The measured point relates to 
the hadronization condition at 158$A$ GeV. 
}
\end{figure}

This would not suggesting that the onset of QGP formation is in the 
collision energy window   40--80$A$ GeV. This would be in conflict with
the result shown in last line of \rt{Tmu}, where we present the 
experimental result for
\begin{equation}\label{Kpi}
\mbox{K}/\pi\equiv\sqrt{\mbox{K}^+/\pi^+\times \mbox{K}^-/\pi^-}.
\end{equation}
The advantage of taking the geometric mean of the charged kaon to 
pion ratios is that the chemical potentials of strangeness and 
baryon number cancel in Boltzmann approximation. The constant value 
we see at SPS leaves no space for any discontinuity in strangeness
production in SPS energy range, including the newly 
available  result at 30 GeV~\cite{Gaz03}. 
This finding contradicts  the claim that the rise of 
$\mbox{K}^+/\pi^+$ ratio with decreasing collision energy suggests
new physics occurring near 30$A$ GeV~\cite{Gaz03a}  (but does not contradict
the fact, which is claimed on false grounds). The
consideration of $\mbox{K}/\pi$ indicates 
that the variation of the baryochemical and strangeness 
chemical potentials  fully account for peak in $\mbox{K}^+/\pi^+$ production. 

The increase in  $\mbox{K}/\pi$ ratio  between RHIC and SPS  energies 
is an indicator of new physics, and an energy scan between
 RHIC and SPS  energies would at first
seem appropriate. However, the 40\% increase we see moving on 
from SPS to RHIC has been in this paper understood as the result of the 
initial state changes and faster, more explosive expansion of the
fireball. Still, it would be good to make sure that this rise is not
step-like, which would suggest a phase change.

Our findings make the issue where is the onset of QGP formation
at first enigmatic. We have seen that specific per baryon
 strangeness and entropy are both evolving smoothly in the 
energy range  beginning at 40$A$ GeV and through the RHIC energies. 
We  see in \rf{PLTMUBLIQ} 
that near 80$A$ GeV supercooling seizes, and below this 
collision energy the particle freeze-out occurs in HG phase 
This means  that if QGP phase is formed below 80$A$ GeV, 
there is reequilibration of particle yields. Still, we
expect that strangeness yield would be preserved, and
any enhancement of strangeness to entropy yield as expressed
by the  K/$\pi$  ratio \req{Kpi} should remain visible. 

The top AGS energy  at 11.6$A$ GeV for the Au beam 
yields~\cite{Ahl99}:
\[
{\mbox{K}/\pi}=0.077.
\] 
An increase of AGS result by 55\% is required to reach the SPS
level present at 30$A$ GeV. This increase is sufficiently sudden to suggest that 
a change in reaction mechanism must  occur between
30 and 10$A$ GeV beam energy. Since the only change we are
expecting is the onset of deconfinement, we have all reason to hope
that a rather sudden rise in K/$\pi$ can be observed in this energy
domain. Naturally we also expect that  evolution of specific strangeness
and entropy production will show a sudden increase in this 
energy domain.

\section{Anything new at LHC?}
\label{lhc}
An often posed question is if 
there can be a future for strangeness as signature of QGP at the LHC
energy scale which is 30 times greater than at RHIC. 
In principle, one would think that the extreme 
conditions expected will reduce the importance of 
strangeness as diagnostic tool, since the difference
between $u,d$ and $s$ flavors will diminish. We do not believe that 
 charm and bottom flavors will `replace' strangeness as an observable,
rather these new degrees of freedom produced in first 
collisions will complement the physics potential of strangeness,
which as we now argue, will further grow at LHC.

The QGP based production processes  at RHIC lead to  a 
highly oversaturated strangeness phase space yield. The mechanisms 
responsible for this are likely to augment as the energy available 
in the nuclear collision increases: it is  probable that  
at LHC  we reach greater
initial temperatures and more explosive transverse flow. As the system 
volume expands, but entropy remains nearly conserved, there is 
significant reduction of particle pairs. 
Since the reannihilation of light quarks is favored by cross
section, compared to the reannihilation of more massive strange quarks, 
this faster expansion is driven 
predominantly by the energy derived from 
light flavors. Strange quarks, 
once produced should remain more if not 
mostly preserved till the hadronization. 

As a consequence,  we expect that $\gamma_s/\gamma_q$ will further increase at 
LHC compared to the large RHIC value, which at $\sqrt{s}=130A$ GeV is about 1.5\,. 
An increase in this ratio triggers an increase in the yield
of kaons, as compared to pions. At the nearly 
baryon free conditions prevailing at LHC, 
we can consider  $\lambda_q=\lambda_s=1$,
hence the value of K$^\pm/\pi^\pm$ ratio in 
a statistical hadronization model is proportional solely to  $\gamma_s/\gamma_q$
and depends on the hadronization temperature $T$. When $\gamma_s=\gamma_q=1$,
this ratio as function of $T$ is shown in \rf{KPIT} as dashed line.
This chemical equilibrium 
result applies also at finite baryon density (SPS)
 to the geometric mean of the 
charged particle ratios, \req{Kpi}.
At $T\simeq 160$ MeV, we see in \rf{KPIT} K$/\pi\simeq 0.133$. 

\begin{figure}[ht]
\vspace*{-4.9cm}
\centerline{\epsfig{width=13.2cm,clip=,figure=\pathnow 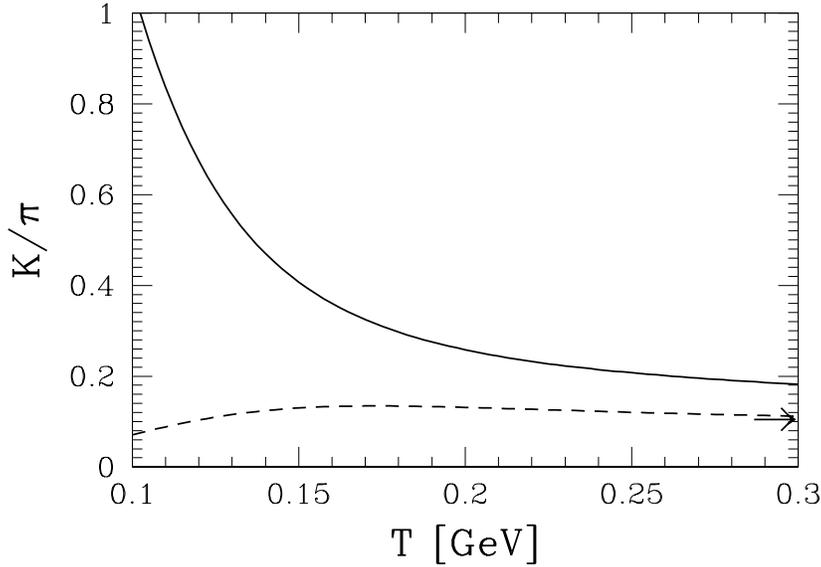}
}
\vspace*{-0.9cm}
\caption{ 
Kaon to pion ratio (see text) as function of 
hadronization temperature for chemical equilibrium (dashed line)
and with maximum allowable  $\gamma_s$ and $\gamma_q$ (solid line).
\label{KPIT} 
}
\end{figure}
 
The solid line, in 
 \rf{KPIT}, shows the maximum possible value of $K/\pi$ ratio, 
with both  $\gamma_s$ and $\gamma_q$
increasing to the maximum allowed by the requirement that neither 
kaon nor pions condensation arises, compare \req{Kcond}: 
\[
\gamma_s\gamma_q=e^{m_{\rm K}/T}, \quad
\gamma_q\gamma_q=e^{m_\pi/T},\ \mbox{for}\ \lambda_q=\lambda_s=1.
\]
The arrow indicates where the lines will meet at 
very large hadronization temperatures. The exact location of this
 limit is in part result of yet
limited knowledge of hadronic high mass resonances. In fact, it can 
be suspected that the slight decrease in the ratio K$/\pi$ with
increasing hadronization temperature is spurious, 
originating from our lack of knowledge
of the hadron strange and non-strange mass spectra. Moreover, it can be expected
that hadronization temperature remains at 
or below  the Hagedorn temperature $T\simeq 160$ MeV at LHC. 

The important result we see is that  K$/\pi$ can almost
triple without violating any fundamental principles, in a scenario
in which initial thermal and  very hot QGP phase is formed and 
expands explosively. Of course, this does not prove that this
will happen at LHC, though a significant increase of  K$/\pi$ 
should  occur. Perhaps more spectacular is the expected rise 
in the Wr\'oblewski ratio $W_s$, which compares the effectiveness
of strange quark production to light quark production~\cite{Wro85}, counting 
only newly made quark pairs, compare~\req{Wro} and~\rf{WSnoe}.
$W_s\simeq 0.2$--$0.25$  in $p$--$p$ reactions 
expressing the low yield of strange quark pairs available. In heavy
ion collisions, values as large as triple this result have been observed, 
expressing, as discussed, the large enhancement of strangeness
yield as compared to light quark yield. Somewhat smaller $W_s$ values 
are expected in chemical equilibrium in baryon-free matter, as shown
by dashed line in  \rf{WST}. The solid line, in \rf{WST},
shows the  great enhancement of $W_s$
should maximal overpopulation of strange quark phase space indeed be 
established (note logarithmic scale in  \rf{WST}).
The arrow indicates the large $T$ limit of $W_s$ which against naive expectation
is found somewhat below unity. This is an expression of
the asymmetry in the number of undiscovered strange and non-strange high mass 
hadron resonances. Again, this is a purely academic point, as hadronization
is expected at or below the Hagedorn temperature $T\simeq 160$ MeV.
\begin{figure}[t]
\vspace*{-4.5cm}
\centerline{\epsfig{width=12.2cm,clip=,figure=\pathnow 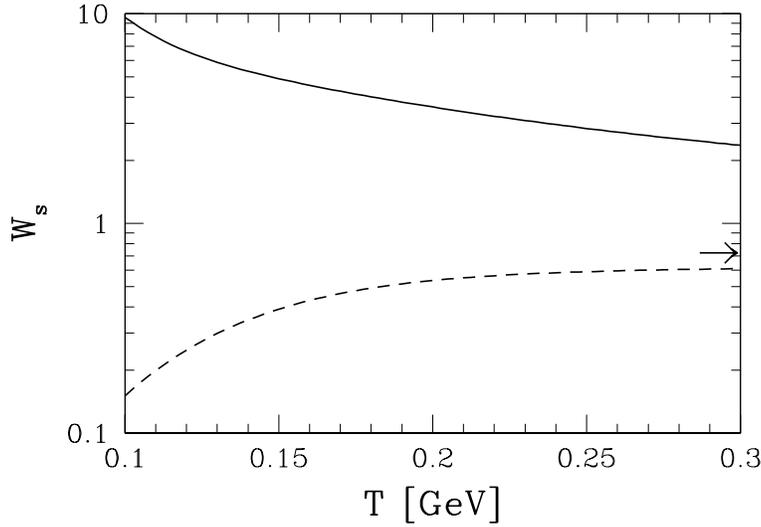}
}
\vspace*{-0.9cm}
\caption{ 
Wr\'oblewski ratio $W_s$  for chemical equilibrium (dashed line)
and maximum allowable  $\gamma_s$ and $\gamma_q$ in baryon-free matter.
\label{WST} 
}
\end{figure}

The new physics  presented in this report, include a consistent
chemical non-equilibrium  analysis of the
SPS and RHIC experimental results which shows great consistency of physical
properties and reaction mechanism in these two domains. We have
demonstrated that the great enhancement of strangeness production
at RHIC is originating in the same mechanism as strangeness production
at SPS. We have 
illustrated how low hadronization temperatures derive from
supercooling due to the fast collective flow of deconfined matter, 
and we have above 
presented a possible scenario for a large strangeness anomaly at LHC. 

\ack
Supported  by a grant from the U.S. Department of
Energy,  DE-FG03-95ER40937\,. Laboratoire de Physique Th\'eorique 
et Hautes Energies, LPTHE, at  University Paris 6 and 7 is supported 
by CNRS as Unit\'e Mixte de Recherche, UMR7589.

\section*{References}


\end{document}